\documentclass[prb,aps,onecolumn,showpacs,floats,a4paper,times,12pt]{revtex4}
\usepackage{epsfig,bm,epsf,graphics}
\begin{document}
\draft
\title{Equation of State from Potts-Percolation Model of a Solid}
\author{Miron Kaufman$^{a}$ and H. T. Diep$^b$\footnote{ Corresponding author, E-mail:diep@u-cergy.fr } }

\address{
$^a$ Department of Physics, Cleveland State University, Cleveland, OH 44115, USA\\
$^b$ Laboratoire de Physique Th\'eorique et Mod\'elisation,
Universit\'e de Cergy-Pontoise, CNRS, UMR 8089\\
2, Avenue Adolphe Chauvin, 95302 Cergy-Pontoise Cedex, France\\
}

\begin{abstract}

We expand the Potts-percolation model of a solid to include stress and strain.  Neighboring atoms are connected by bonds.  We set the energy of a bond to be given by the Lennard-Jones potential.  If the energy is larger than a threshold the bond is more likely to fail, while if the energy is lower than the threshold the bond is more likely to be alive.  In two dimensions we compute the equation of state: stress as function of inter-atomic distance and temperature by using renormalization group and Monte Carlo simulations. The phase diagram, the equation of state, and the isothermal modulus are determined.  When the Potts heat capacity is divergent the continuous transition is replaced by a weak first-order transition through the van der Waals loop mechanism.  When the Potts transition is first order the stress exhibits a large discontinuity as function of the inter-atomic distance.
\end{abstract}

\maketitle
\section{Introduction}

	The mechanical properties of solids, such as melting\cite{Gomez2001,Gomez2003} and mechanical failure\cite{Arcangelis,Beale,Wang,Alava}, are topics of considerable interest. In this paper we continue the analysis of an equilibrium statistical mechanics model\cite{Englman,Blumberg} of a solid. Previously we assumed\cite{Kaufman2008} harmonic springs and evaluated the role of thermal fluctuations by using renormalization-group and Monte-Carlo simulations.  Furthermore we studied\cite{Diep2009} the model with an extended defect line and found a hybrid, first- and second-order phase transition.  In this paper we study the equation of state of the solid, stress as function of strain and temperature, obtained by assuming neighboring atoms are separated by a fixed inter-atomic distance and associating to each pair of neighboring atoms the Lennard-Jones energy.
An alternative realistic anharmonic energy versus atomic distance due to Ferrante\cite{Ferrante}  has been considered
in a previous work\cite{Kaufman96}.

	The model is defined in Section II.  We assume the energy of a pair of neighboring atoms to be given
by the Lennard-Jones 6-12 potential.  If the energy of such a spring is larger than the threshold energy,
the probability for its failure is higher than 50\%.  This model is mapped into a Potts model with couplings
that are dependent on the inter-atomic distance.  The free energy, number of live bonds,
their fluctuations, stress, and modulus are computed using renormalization group and
Monte Carlo techniques.  Fixing the inter-atomic distance to a value independent of the atoms locations makes this model mean-field like. The role of the inter-atomic distance fluctuations will be studied in a future work. The number of Potts states q is a fugacity conjugated to the number of clusters of live bonds, with $q = 1$ corresponding to springs failing independently of each other.  A challenging question not addressed here is how to connect $q$ to data from a real solid.

	The equation of state is studied paying particular attention to the solid failure
signaled by an extremum in the dependence of stress on the inter-atomic distance.  Beyond the maximum under
expansion and the minimum under compression for stress versus strain dependence, the solid ceases to be thermodynamically stable. The phase diagram includes the Potts transition line.  If the Potts transition line is in the stable region of the phase diagram and if the $q$ value is such that the Potts heat capacity is divergent, a remarkable phenomenon occurs in the vicinity of the transition line.  A van der Waals loop\cite{Waals} developes in the stress-strain dependence signaling a weak phase transition that replaces the continuous Potts phase transition. While in the renormalization group calculations\cite{Migdal, Kadanoff} that are exact\cite{Berker} on hierarchical lattices\cite{Griffiths82,Kaufman84}the Potts transitions are always continuous, in the Monte Carlo simulations we can see both continuous transitions (for small $q$) and discontinuous transitions (for large $q$). This allows us to explore the influence of the order of the Potts transition on the mechanical properties of the solid, such as stress dependence on temperature and interatomic distance.

	In Section III we present numerical results based on the renormalization-group Migdal-Kadanoff scheme.
Monte-Carlo simulations are presented in Section IV.  Our concluding remarks are found in Section V.

\section{Model }

The energy of any pair of neighboring atoms is:

\begin{equation}\label{LJ}
E(r)=\epsilon [(\frac {r_0}{r})^{12}-2 (\frac {r_0}{r})^{6}]
\end{equation} 				
where $r$ is the inter-atomic distance and $r_0$ is the equilibrium inter-atomic distance (under zero stress).  If the energy of the spring is larger than the threshold energy $E_0$ the bond is more likely to fail than to be alive. $p$ is the probability that the bond is alive and $1- p$ the probability that the bond is broken.  We assume the probabilistic weight $w = p/(1-p)$ to depend on energy through the Boltzmann weight:

\begin{equation}\label{w}
w=\frac{p}{1-p}=e^{-\frac {E(r)-E_0}{k_BT}}
\end{equation}

We allow for correlations between failing events by using the Potts number of states $q$, which plays the role of a fugacity controlling the number of clusters.  For the same number of live bonds, graphs with more clusters are favored if $q > 1$, while if $q << 1$ there is a tendency to form a few large clusters. If $q = 1$ bonds fail independently of one another, i.e. random percolation process.

The partition function is obtained\cite{Kaufman84a} by summing over all possible configurations of bonds arranged on the lattice

\begin{equation}\label{part}
Z=\sum_{config}q^Cw^B
\end{equation}  					
$C$ is the number of clusters, including single site clusters, and $B$ is number of live bonds.
	
The free energy per bond is $f = \ln Z/N_{bonds}$.  The derivatives of the free energy $f$ with respect to $w$ and $q$ provide respectively the number of live bonds $b$ and the number of clusters $c$, each normalized by the total number of lattice bonds
\begin{eqnarray}
b&=&w\frac{\partial f}{\partial w}\\
c&=&q\frac{\partial f}{\partial q}
\end{eqnarray}

The derivatives of b and c with respect to w and q provide the fluctuations (variances) of those quantities:
\begin{eqnarray}
\Delta b^2&=&w\frac{\partial b}{\partial w}\\
\Delta c^2&=&q\frac{\partial c}{\partial q}
\end{eqnarray} 						

	The stress $\sigma$ is calculated by taking the derivative of the free energy with respect to $r$, the inter-atomic distance:

\begin{equation}\label{eqst}
\sigma=-k_BT \frac{\partial f}{\partial r}= b \frac{dE(r)}{dr}
\end{equation}
 					
The equation of state, Eq. (\ref{eqst}), states that the average stress is equal to the stress associated with each live bond $dE(r)/dr$ multiplied by the number of live bonds $b$.  The energy gradient is obtained from Eq. (\ref{LJ}):

\begin{equation}\label{LJd}
\frac{dE(r)}{dr}=-\frac{12\epsilon}{r_0} \left[(\frac{r_0}{r})^{13}- (\frac{r_0}{r})^{7}\right]
\end{equation} 					

	The equation of state yields the isothermal linear modulus (inverse compressibility) $m$:

\begin{eqnarray}
m&=&\frac{\partial \sigma}{\partial r}|_T=\frac {d^2E(r)}{dr^2}-\frac {1}{k_BT}\Delta b^2 (\frac {dE(r)}{dr})^2\\
\end{eqnarray}
 		
We perform Monte Carlo simulations of the model using its mapping into the Potts model.  By using the Kasteleyn-Fortuin expansion\cite{Wu,Fortuin} we can rewrite the partition function, Eq. (\ref{part}), as:

\begin{equation}\label{part2}
Z=Tr_{\sigma} e^{-\frac {H}{k_BT}}
\end{equation}

The Hamiltonian is:

\begin{equation}\label{hamil}
-\frac{H}{k_BT}=\sum_{<i,j>} J(r)\delta (s_i,s_j)
\end{equation}
where $s_i$ is a Potts\cite{Potts} spin at the lattice site $i$ taking $q$ values.  The coupling constant $J(r)$ is related to the original parameters by

\begin{equation}\label{interact}
J(r)=\ln (1+w)=\ln (1+e^{-\frac {E(r)-E_0}{k_BT}})
\end{equation}

To get the equation of state from Monte Carlo simulations for a given $T$, $r$, and $E_0$, we calculate the Potts coupling constant $J(r)$, using Eq. (\ref{interact}).  Then we rewrite Eq. (\ref{eqst}) to get the stress. The number of live bonds $b$ is obtained from the Potts energy
\begin{equation}\label{u}
u=-\overline {\delta (s_i,s_j)}
\end{equation}
The equation of state becomes
\begin{equation}\label{eqst1}
\sigma=-u[ e^{\frac {E(r)-E_0}{k_BT}}+1 ]^{-1}\frac {dE(r)}{dr}
\end{equation}
 				
	In all numerical results that will be presented next in Section III from renormalization group calculations and in Section IV from Monte Carlo simulations, we express energy in units of $\epsilon$, temperature in units of  $\epsilon/k_B$, distance in units of $r_0$ and stress in units of $\epsilon/r_0$.

\section{ Renormalization Group}

The Migdal-Kadanoff\cite{Migdal, Kadanoff} recursion equation for two dimensions is

\begin{equation}\label{RG1}
w'=[ 1+ \frac{w^2}{2w+q} ]^{2}-1
\end{equation}

The free energy $f = \ln Z/N_B$, $N_B$ being the number of lattice edges, is
\begin{equation}\label{RG2}
f=\sum_N \frac{C_N}{4^N}
\end{equation} 						
where
\begin{equation}\label{RG3}
C=2\ln (2w+q)
\end{equation} 	

The recursion equations (\ref{RG1})-(\ref{RG3}) represent the exact solutions\cite{Berker} for the diamond
hierarchical lattice\cite{Griffiths82,Kaufman84}.

The renormalization group flows are governed by the following fixed points: i. $w = 0$ (non-percolating live bonds),
ii. $w = \infty$ (percolating network of live bonds), iii. $w = w_c$ (Potts critical point).
Using the free energy we can compute the number of live bonds $b$, their fluctuation (variance), stress, thermal expansion and modulus. Each of those quantities is scaled by the total number of lattice bonds.

	For given values of $E_0$ and $q$, the phase diagram in the plane $(T, r)$ includes two type of singularity lines. In Fig. 1 we fixed the threshold energy at $E_0=-0.5$ and the clusters fugacity at $q = 10$.  The Potts- percolation line separates the region where the probability for formation of an infinite cluster of live bonds is non-zero from the region where the probability is zero.  On this line the probability weight defined in Eq. (\ref{w}) is equal to $w_c$.  On the instability line the solid becomes soft, as its modulus vanishes $\partial \sigma/\partial r =0$.  The minimum of stress versus inter-atomic distance under compression ($r < 1$) and its maximum under expansion ($r > 1$) provide the limits of stability for the solid.  The solid becomes soft at these points.  Beyond this line the solid is thermodynamically unstable since $\partial \sigma/\partial r < 0$. Note that a large portion of the Potts transition line (solid,red) is situated in the unstable region.

\begin{figure}
\centering
\includegraphics[width=5.0in]{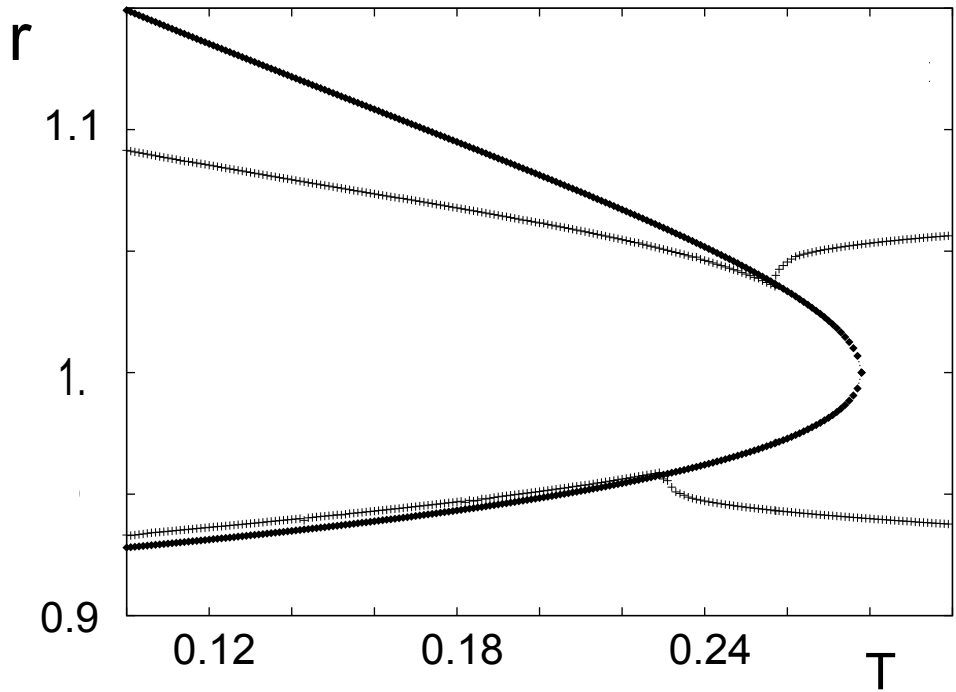}
\caption{ Phase diagram $(T, r)$ plane for $E_0 = -0.5$, $q = 10$. Line with crosses is instability line and line with diamonds is Potts transition line.  The model solid is thermodynamically unstable outside the instability line.} \label{fig:1}
\end{figure}

We show in Fig. 2 the stress versus inter-atomic distance for two isotherms $T = 0.2$ and $T = 0.26$ respectively.  Fig.3 contains zooms close to the Potts transitions that show the van der Waals loops. The modulus is shown in Fig. 4 for the same values of the model parameters.

\begin{figure}
\centering
\includegraphics[width=5in]{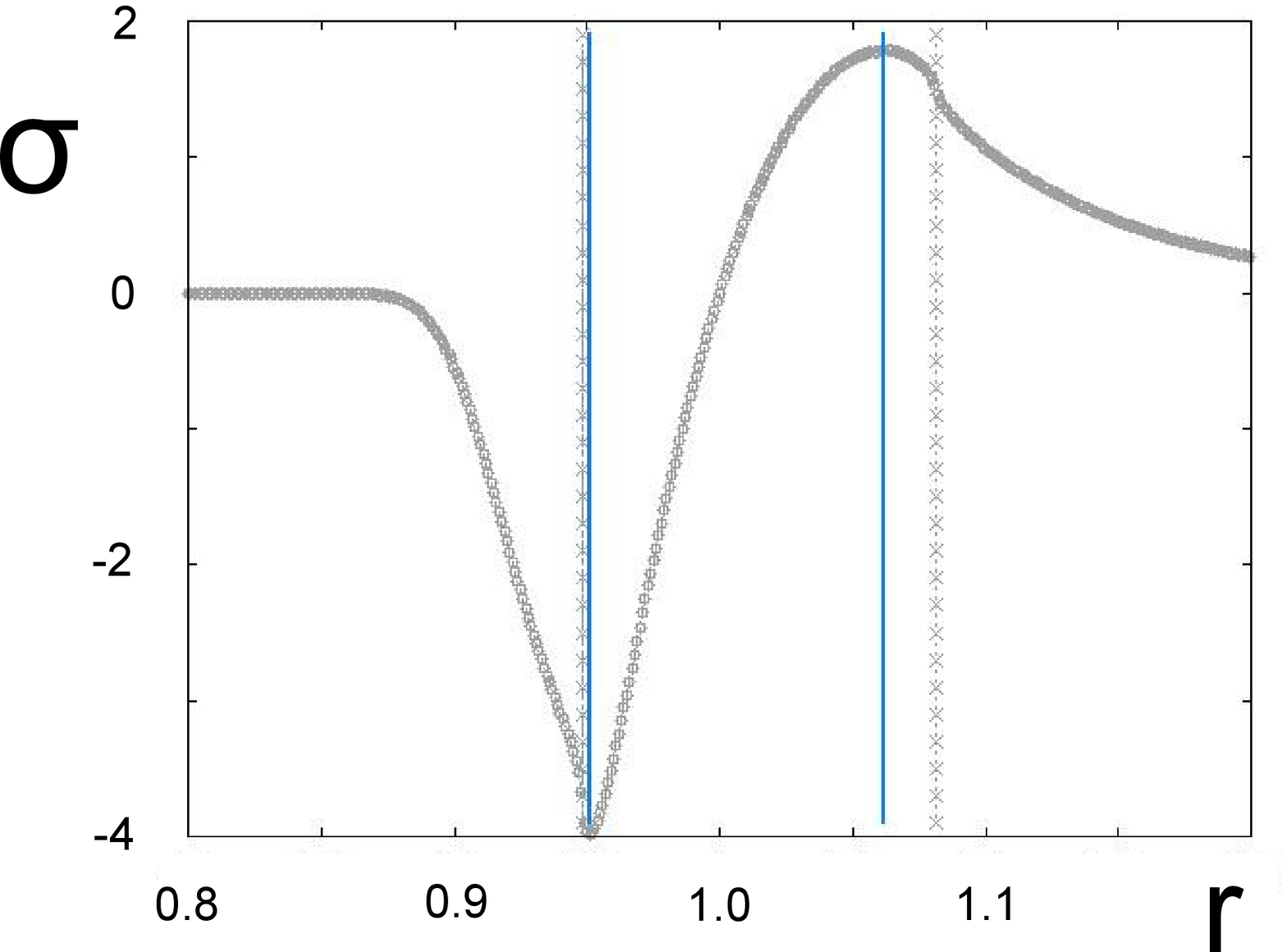}
\includegraphics[width=5in]{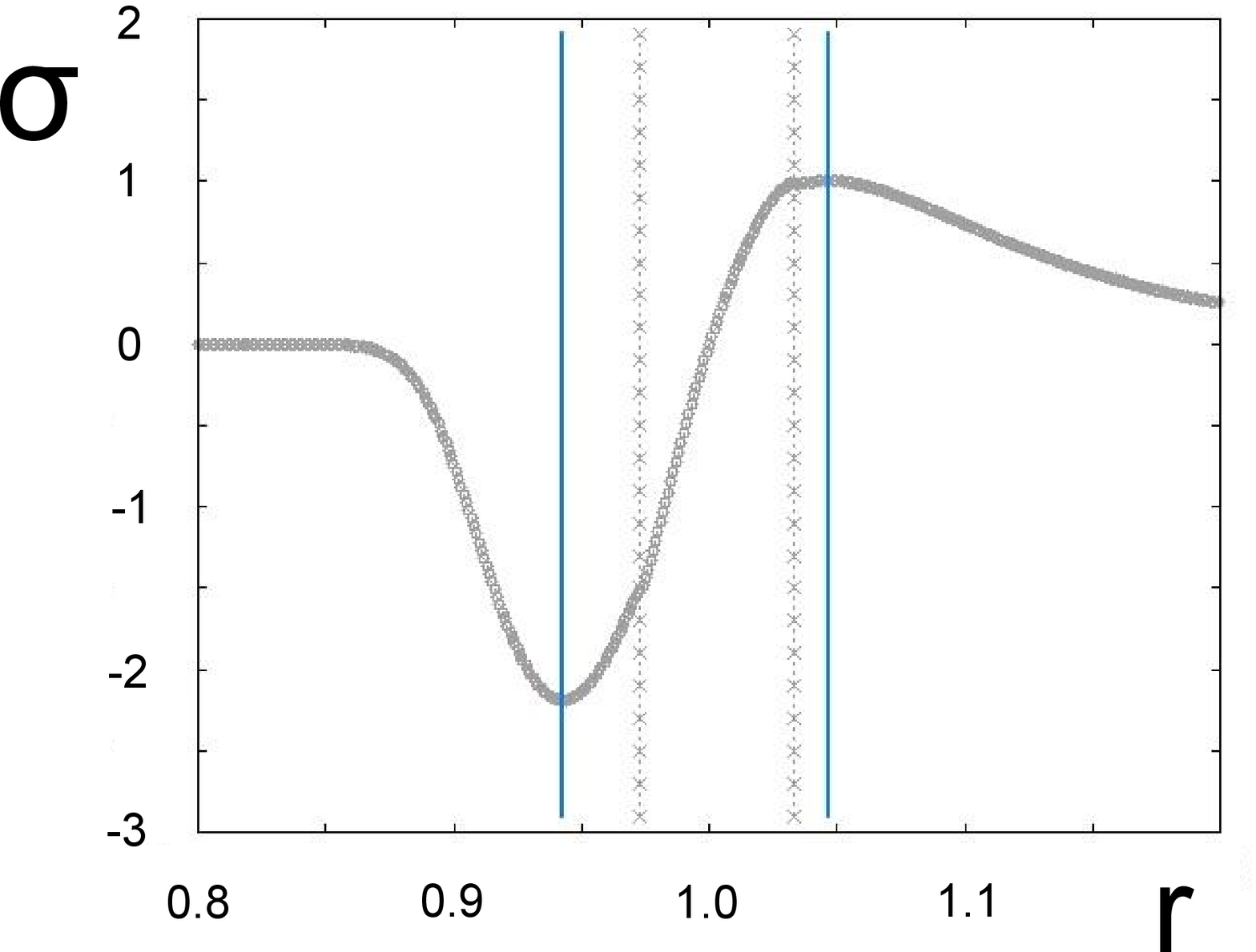}
\caption{ Equation of state isotherms: stress (white circles) versus strain for $E_0 = -0.5$, $q = 10$, $T = 0.2$ (upper) and 0.26 (lower).  The Potts critical points are indicated by the  lines with crosses and the stability limits are indicated by the solid vertical lines. The region between the solid lines is the stable region. The compression region to its left and the expansion region to its right are unstable.. In Fig. 3 we show zooms close to the Potts critical points.
} \label{fig:2}
\end{figure}

The $T = 0.2$ isotherm is in the region of phase diagram where the Potts transition is in the instability region while the $T =0.26$ one is in the region where the Potts transition is in the stable region.  A zoom view on the isotherm $T = 0.26$ in the neighborhoods of the two Potts transitions (see Fig. 3) reveals van der Waals loops\cite{Waals} that yield weak discontinuous transitions (small discontinuity) by means of the Maxwell construction.  Hence discontinuous transitions replace the continuous transition of the Potts model for the hierarchical lattice.  The modulus and the thermal expansion (Fig. 4) show anomalies at the Potts transition that are connected to the divergence of $\Delta b^2$ (or of the heat capacity) in the Potts model on the diamond hierarchical lattice for $q = 10$.  The modulus becomes negative, and thus the solid is thermodynamically unstable, when $\Delta b^2$ is large enough, see Eq. (8). For the square lattice we expect the same phenomenon for $q = 2, 3, 4$ where the heat capacity (and thus $\Delta b^2$) is infinite at the Potts critical point.  This will be confirmed through Monte Carlo simulations in Sec IV below.
 	
\begin{figure}
\centering
\includegraphics[width=5in]{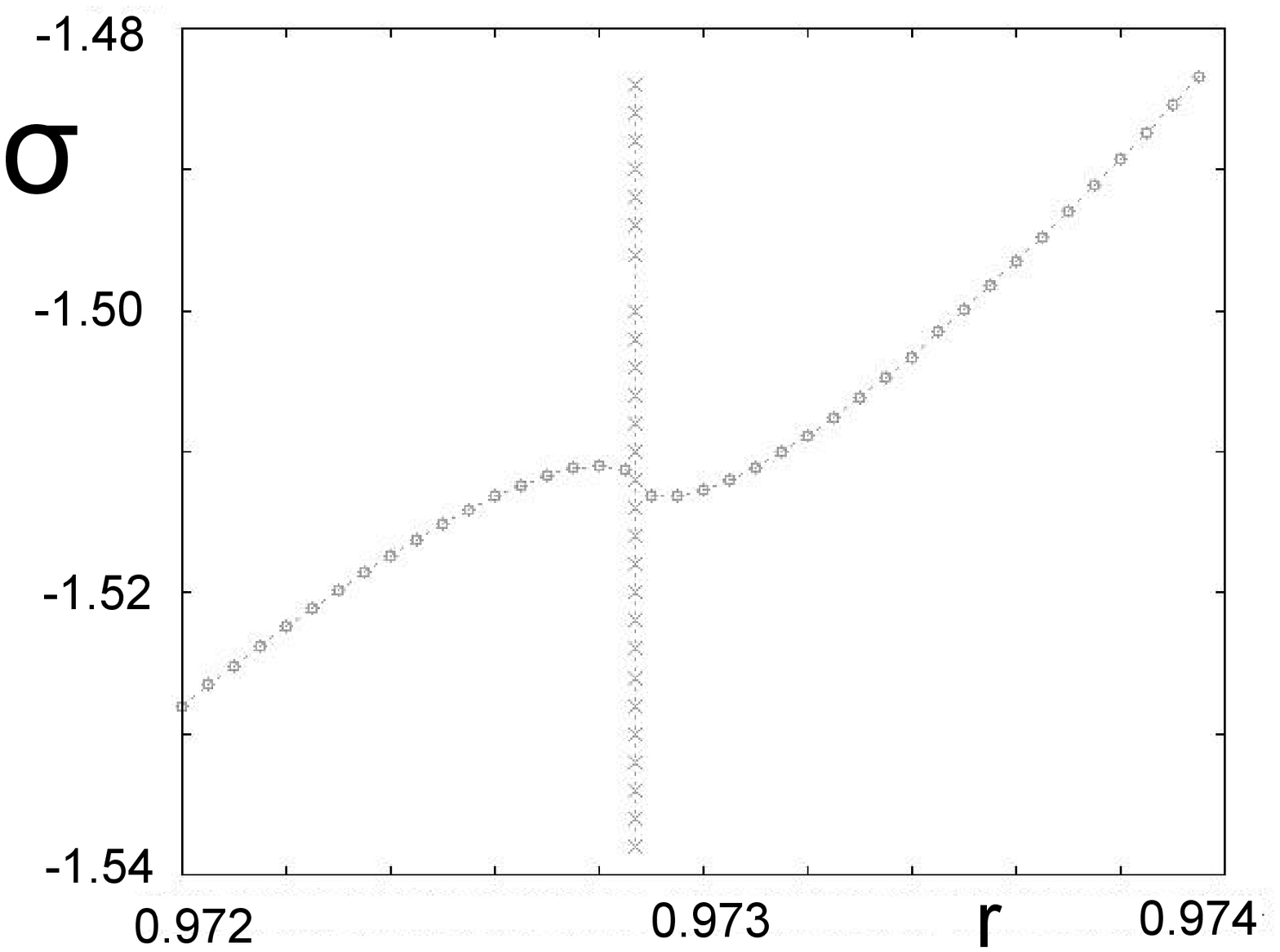}
\includegraphics[width=5in]{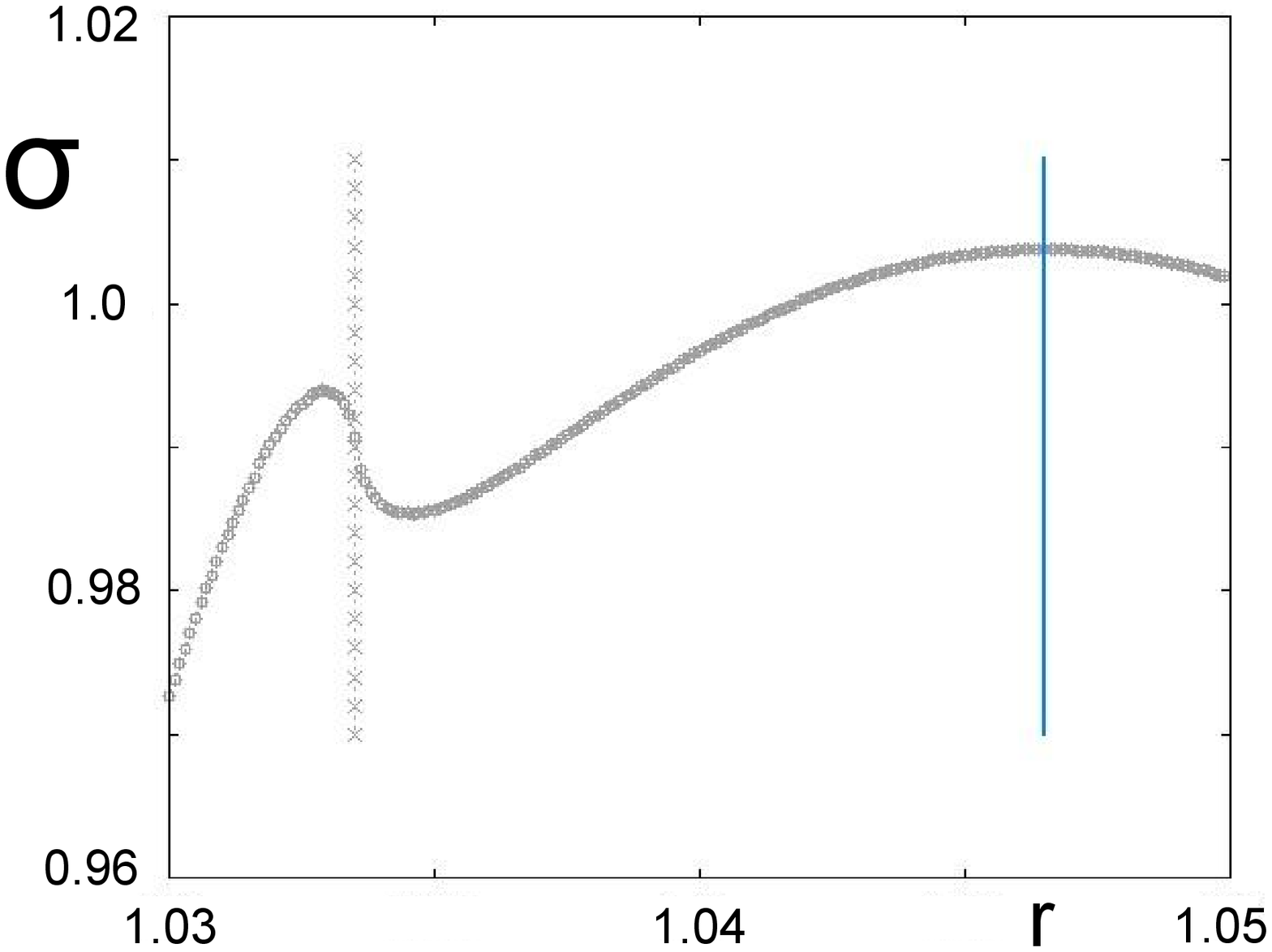}
\caption{ Close-ups on isotherm $E_0 = -0.5$, $q = 10$, $T = 0.26$ near the Potts transition line under compression and expansion, respectively. van der Waals loops, that signal discontinuous transitions through the Maxwell construction, are exhibited at the Potts critical points (indicated by vertical lines with crosses). The solid line shows the stability limit.} \label{fig:3}
\end{figure}

\begin{figure}
\centering
\includegraphics[width=5in]{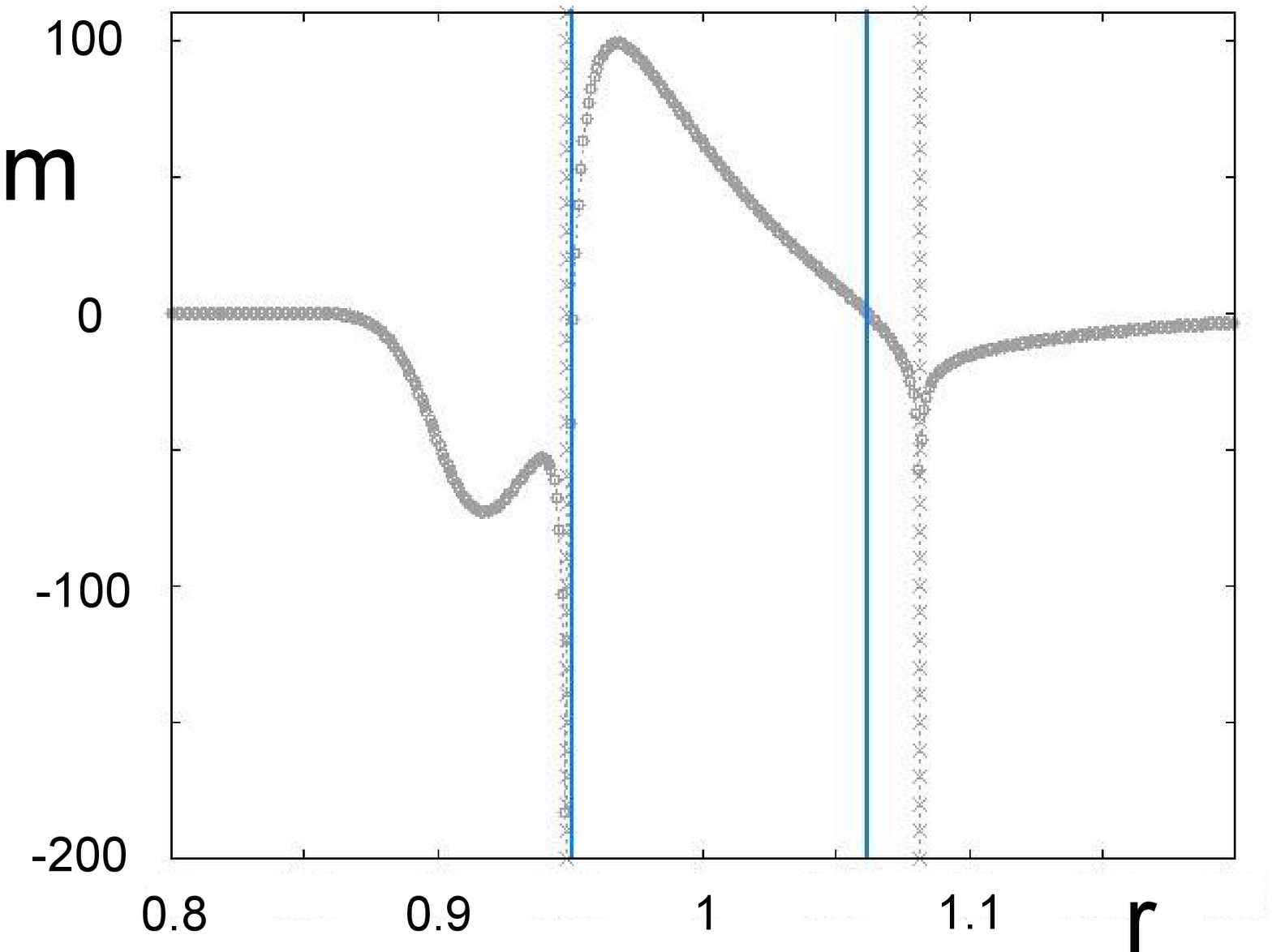}
\includegraphics[width=5in]{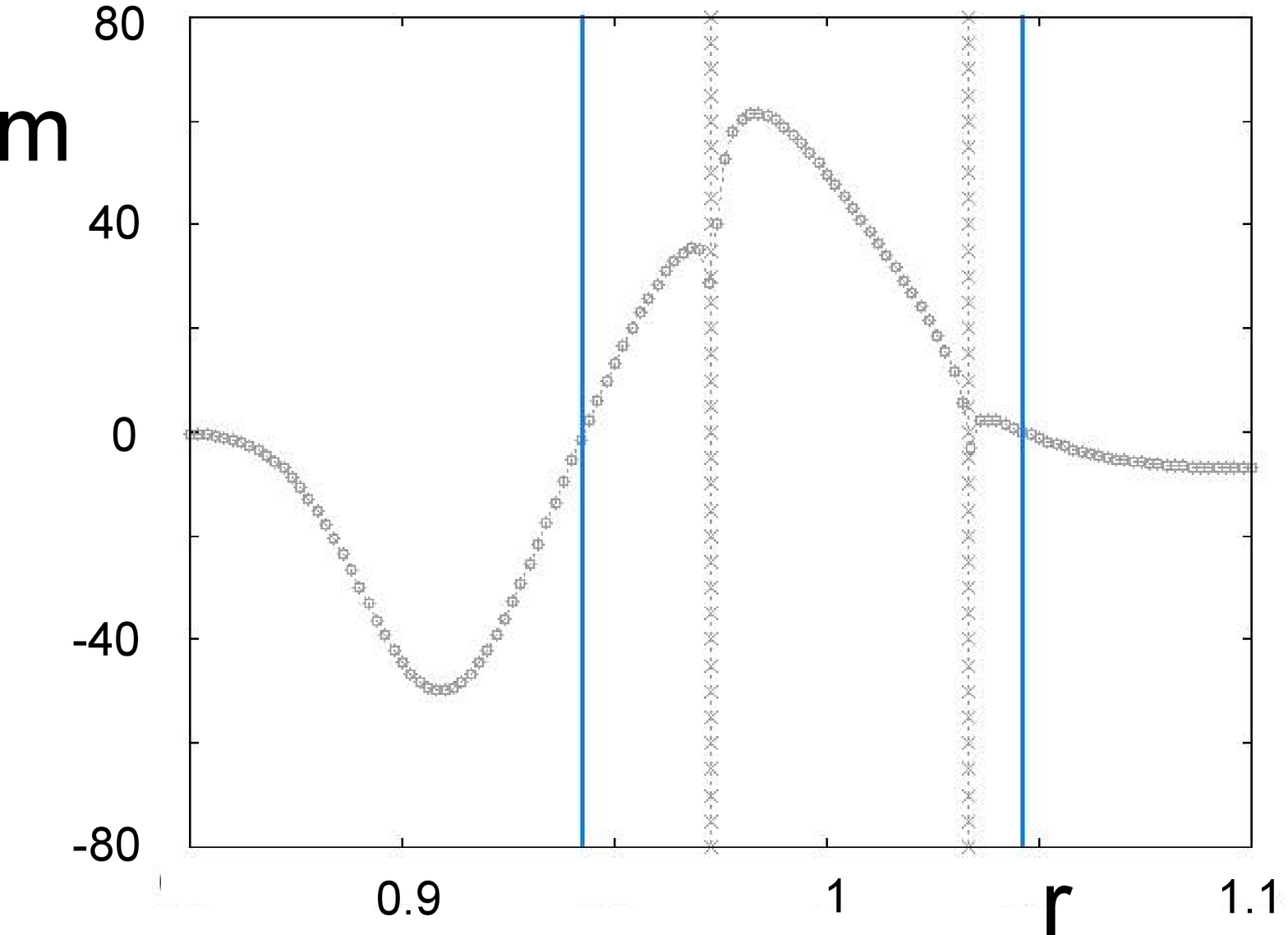}
\caption{ Isothermal modulus $m$ versus inter-atomic distance $r$ for $E_0 = -0.5$, $q = 10$, $T = 0.2$ (upper) and 0.26 (lower).  The Potts transitions are indicated by the vertical lines with crosses, while the solid lines show the stability limits. The region between the solid lines is the stable region. The compression region to its left and the expansion region to its right are unstable..} \label{fig:4}
\end{figure}

In Fig. 5 we show the isotherm where the temperature equals the critical temperature for $r = 1$ (equilibrium inter-atomic distance for zero stress).  We also show the corresponding modulus.  The anomaly apparent in Figs. 2-4 is not present at $r = 1$ because the derivative of $E$ with respect to $r$  [see Eq. (9)] is zero at $r = 1$ and thus the negative contribution on the right hand side of Eq. (10) vanishes.

\begin{figure}
\centering
\includegraphics[width=5in]{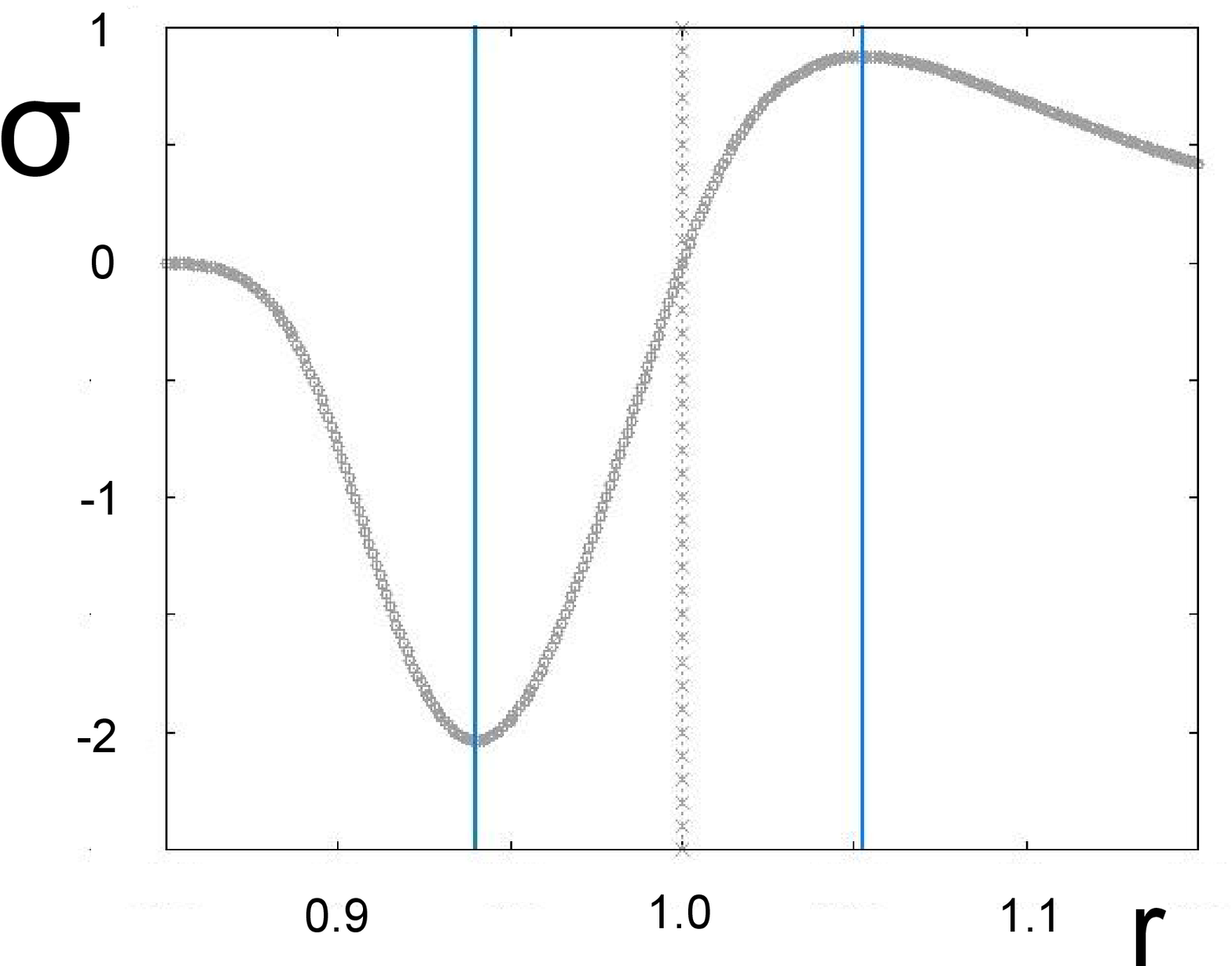}
\includegraphics[width=5in]{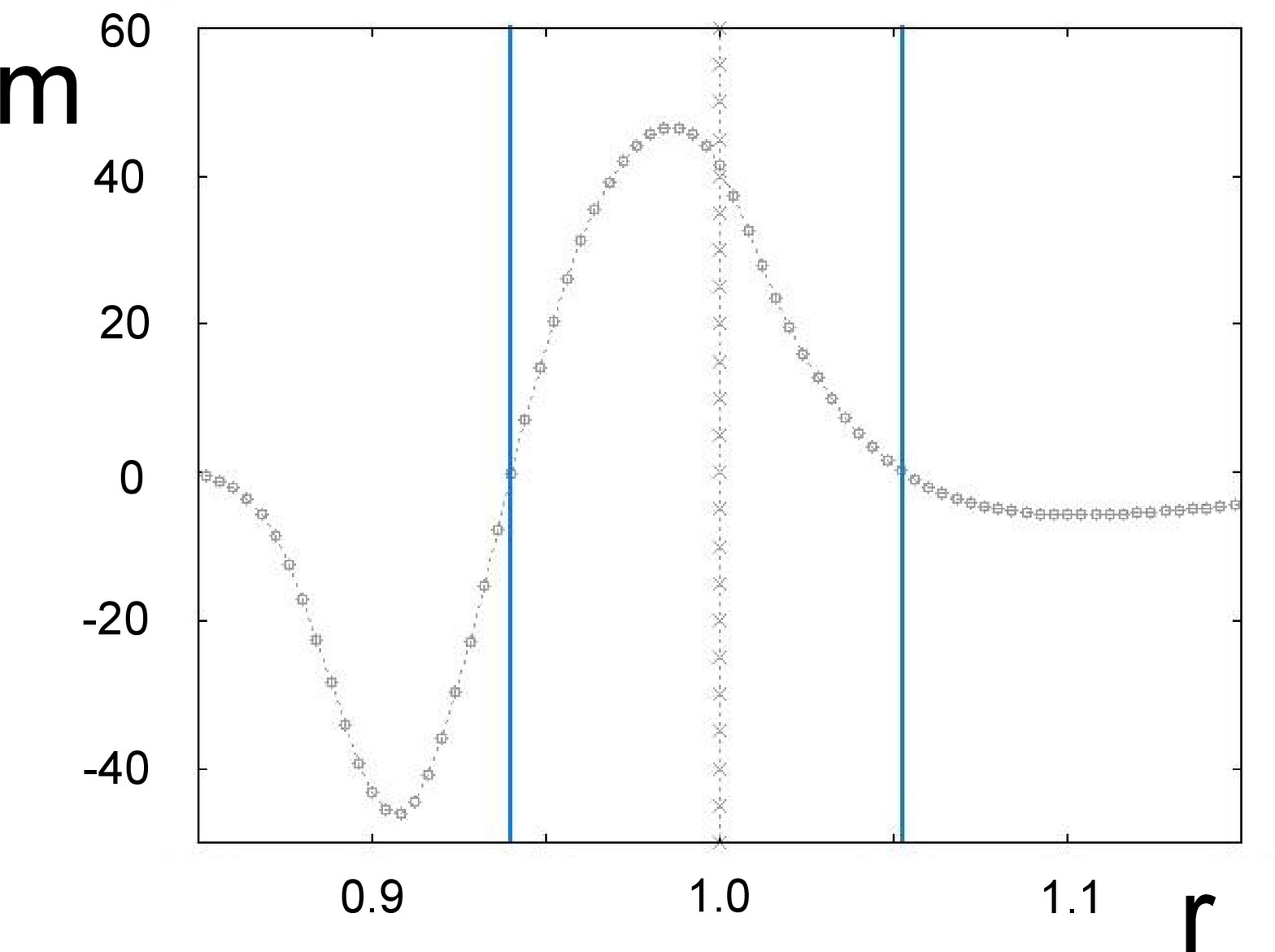}
\caption{ Stress and modulus versus inter-atomic distance at the critical temperature corresponding to $r = 1$.  The Potts critical point is indicated by the line with crosses. No anomaly occurs because of the vanishing of $\frac {dE}{dr}$.  The region between the solid lines is the stable region. The compression region to its left and the expansion region to its right are unstable..} \label{fig:5}
\end{figure}

	To illustrate the dependence of the phase diagram on the parameters $q$ and $E_0$ we show, beside the phase diagram for $E_0 = -0.5$, $q = 10$ of  Fig. 1, the phase diagram for $E_0 = 0.5$, $q = 10$ in Fig. 6, and the phase diagrams for $E_0 = -0.5$, $q = 1$, and for $E_0 = 0.5$, $q = 1$ in Fig. 7.  The Potts line in the $(T,r)$ plane is obtained by substituting in Eq. (2) the critical value $w_c(q)$, $r_0 = 1$ and using Eq. (1):

\begin{equation}\label{critline}
(\frac {1}{r})^{12}-2 (\frac {1}{r})^{6}=E_0-T\ln(w_c(q))
\end{equation}

For the self-dual square lattice the critical value\cite{Wu} is $w_c(q) = \sqrt {q}$ and this will be used in Sec.IV below to verify the accuracy of the Monte Carlo simulations. 	
The instability line originates at zero temperature at the two values of r for which $E(r) = E_0$ provided the the threshold energy is $-1 < E_0 < -0.787$. The energy value of -0.787 is the energy at the inflexion point $r = 1.11$.   If the threshold energy $E_0 > -0.787$ the instability line in the expansion region $r > 1$ originates from the inflexion point of $E(r)$, while under compression the starting point is at the $r$ for which $E(r) = E_0$.

\begin{figure}
\centering
\includegraphics[width=5.0in]{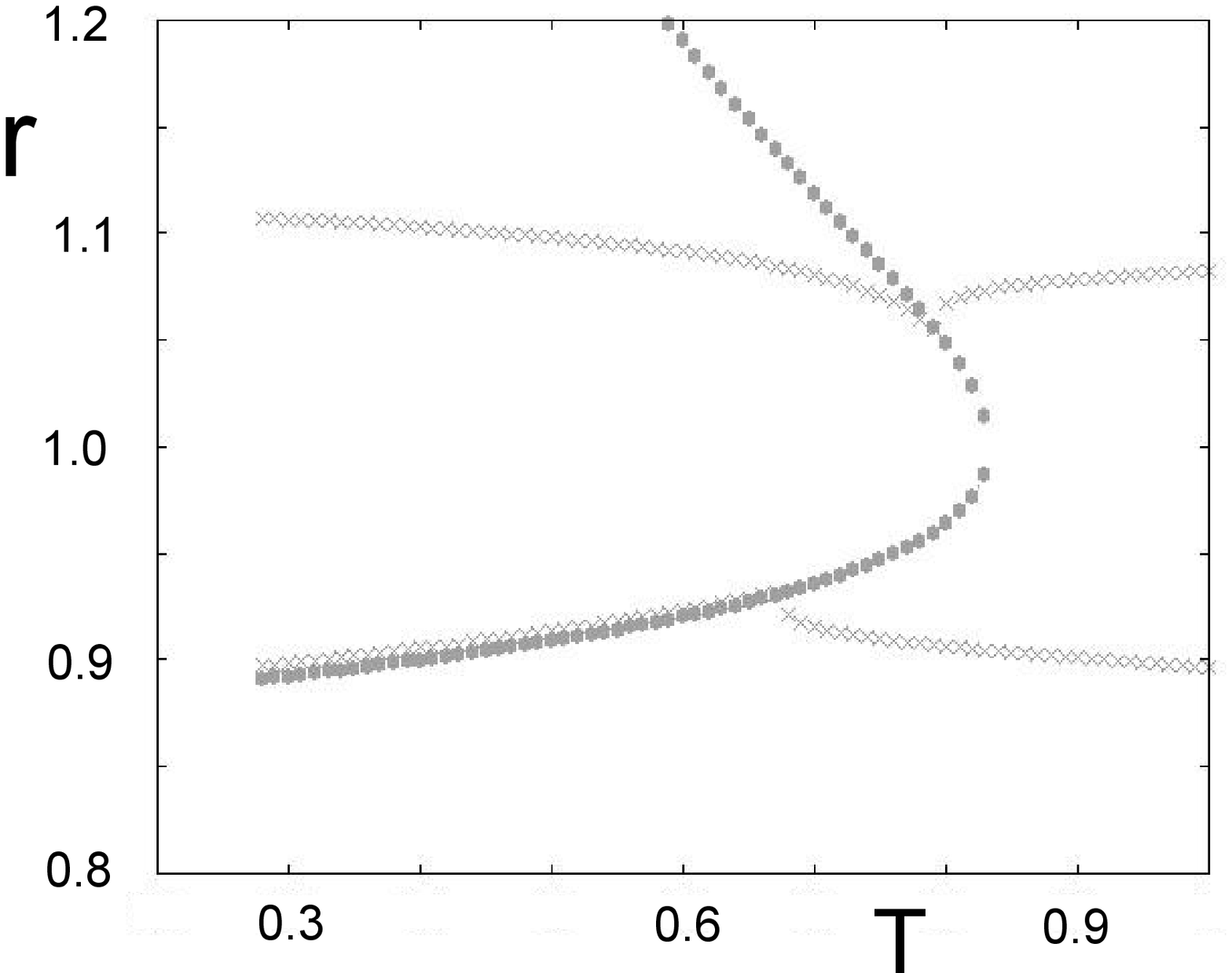}
\caption{Phase diagram $(T, r)$ plane for $E_0 = 0.5$, $q = 10$. Lines displayed by crosses are instability lines and the line with black circles is the Potts transition line. The model solid is thermodynamically unstable outside the instability lines.} \label{fig:6}
\end{figure}

\begin{figure}
\centering
\includegraphics[width=5.0in]{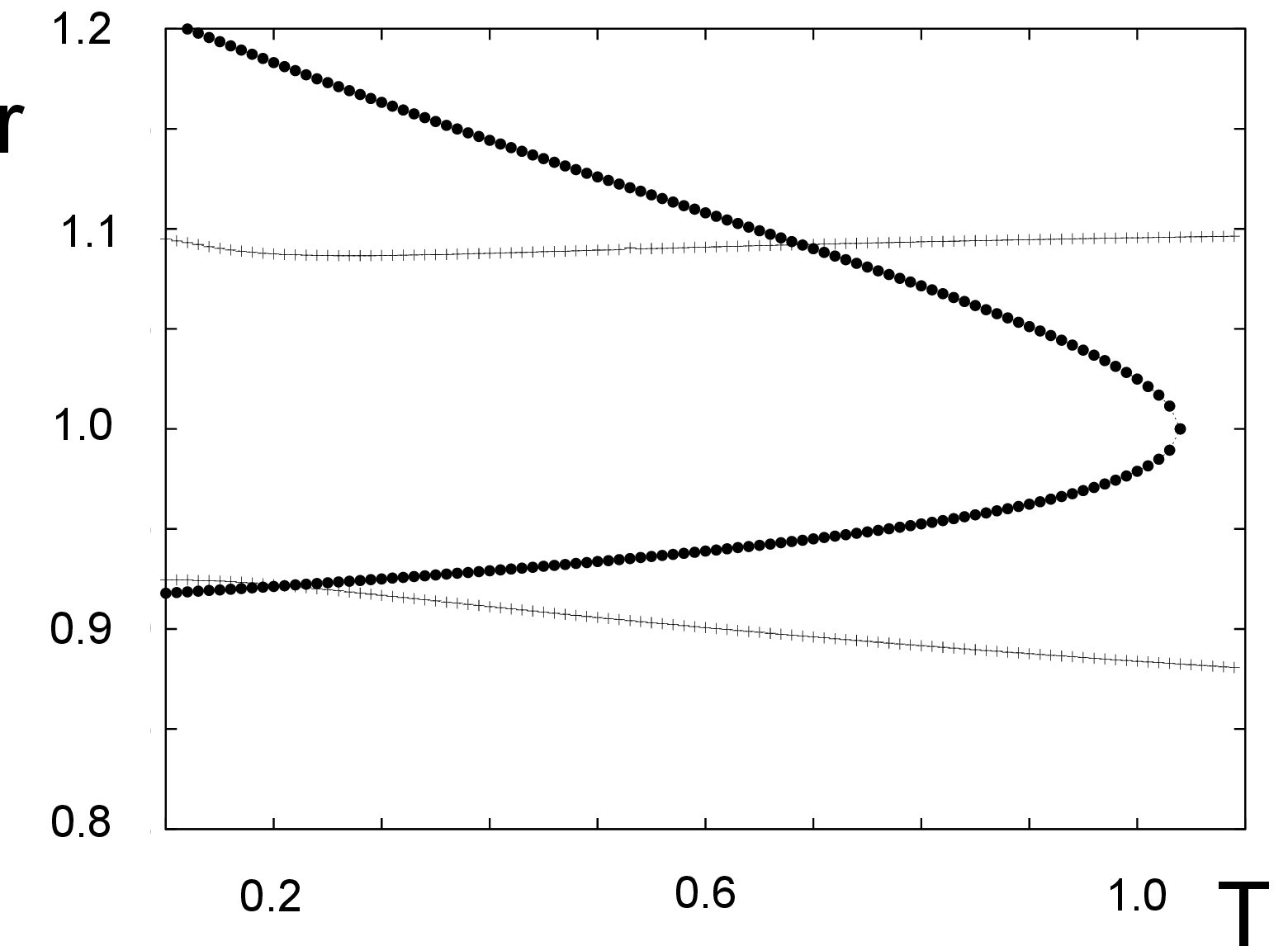}
\includegraphics[width=5.0in]{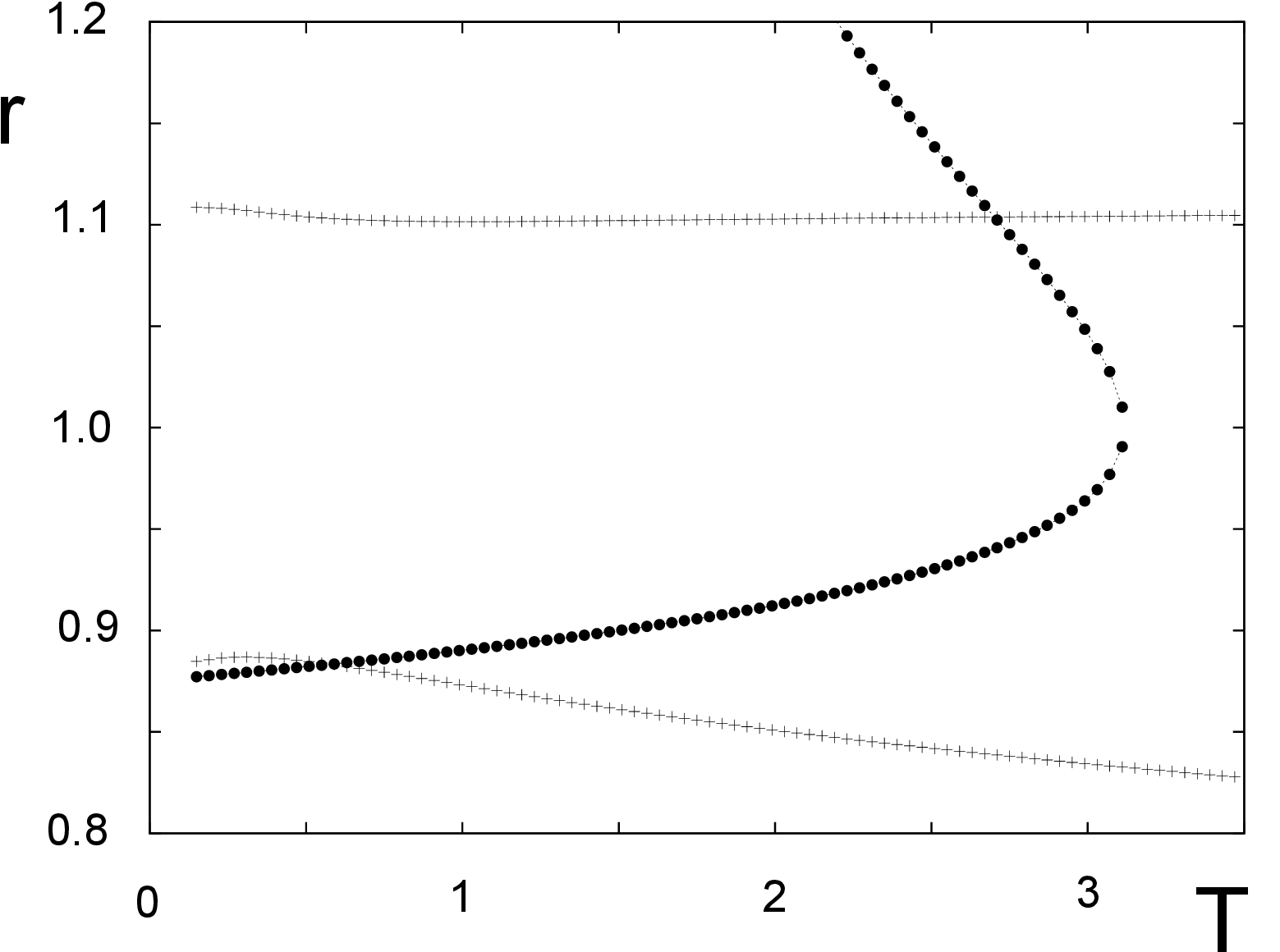}
\caption{Phase diagram $(T, r)$ plane for $E_0 = -0.5$ (upper) and $E_0 = 0.5$ (lower) for $q = 1$. Lines displayed by crosses are instability lines and  line with black circles is Potts transition line. The model solid is thermodynamically unstable outside the instability lines.} \label{fig:7}
\end{figure}


\section{Monte Carlo Simulations}

\subsection{General remarks}

We perform here Monte Carlo simulations of the Potts model governed by the Hamiltonian in Eq. (\ref{hamil}) with the nearest-neighbor interaction $J(r)$ given by   Eq. (\ref{interact}).   For a given $q$, the main parameters are $T$, $r$ and $E_0$.  As in the previous section, we fix $q$ and $E_0$ and make simulations in the space $(T,r)$.

The details of the simulations are the following. We consider a square lattice of
size $N_x\times N_y$ where $N_x=N_y=40, 60, 80,100$.  Each lattice site is
occupied by a $q-$state Potts spin.   We use periodic boundary
conditions. Depending on the location of the studied point in the phase space, we used an equilibrating time from $10^5$ to $10^6$ MC steps per spin and an averaging time of the order of $10^6$ MC steps per spin.

One purpose here is to test the renormalization group prediction of section III for the cases when the heat capacity of the Potts model is divergent at criticality: $q = 2, 3, 4$.  Though the renormalization group analysis is exact for diamond hierarchical lattice, its predictions for the square lattice are to be checked.  The other goal of the simulations is to learn about the influence of the discontinuous Potts transition for $q > 4$ on the equation of state of our model. To achieve this we simulate the $q = 10$ on the square lattice.

To verify the accuracy of our simulations we compared the critical temperature estimated from our simulations to the exact critical temperature. For $r=1$ it is:
\begin{equation}
T_c=\frac {1+E_0}{\ln (\sqrt{q})}
\end{equation}
With $q=4$, $E_0=-0.5$, we find $T_c\simeq 0.721$. The Monte Carlo simulations give the same result up to 4 digits.

In MC simulations, we work at finite sizes, so for each size we have to determine the "pseudo" transition which corresponds in general to the maximum of the specific heat or of the susceptibility. The
maxima of these quantities need not to be at the same temperature. Only at the infinite size, they should coincide. The theory of finite-size scaling permits to deduce properties of a system at its thermodynamic limit.  We have used in this work a size large enough to reproduce the bulk transition temperature up to the fourth decimal.  We define the Potts order parameter $Q$ by
\begin{equation}\label{Q}
Q=[q\max (Q_1,Q_2,...,Q_q)-1]/(q-1)
\end{equation}
where $Q_n(n=1,...,q)$ is the averaged value defined by
\begin{equation}\label{Qn}
Q_n=<\sum_j \delta (s_j-n)/(N_xN_y)>
\end{equation}
$s_j$ being the Potts spin at the site $j$.

 \subsection{Results for q=10}

\begin{figure}
\centering
\includegraphics[width=2.5in,angle=0]{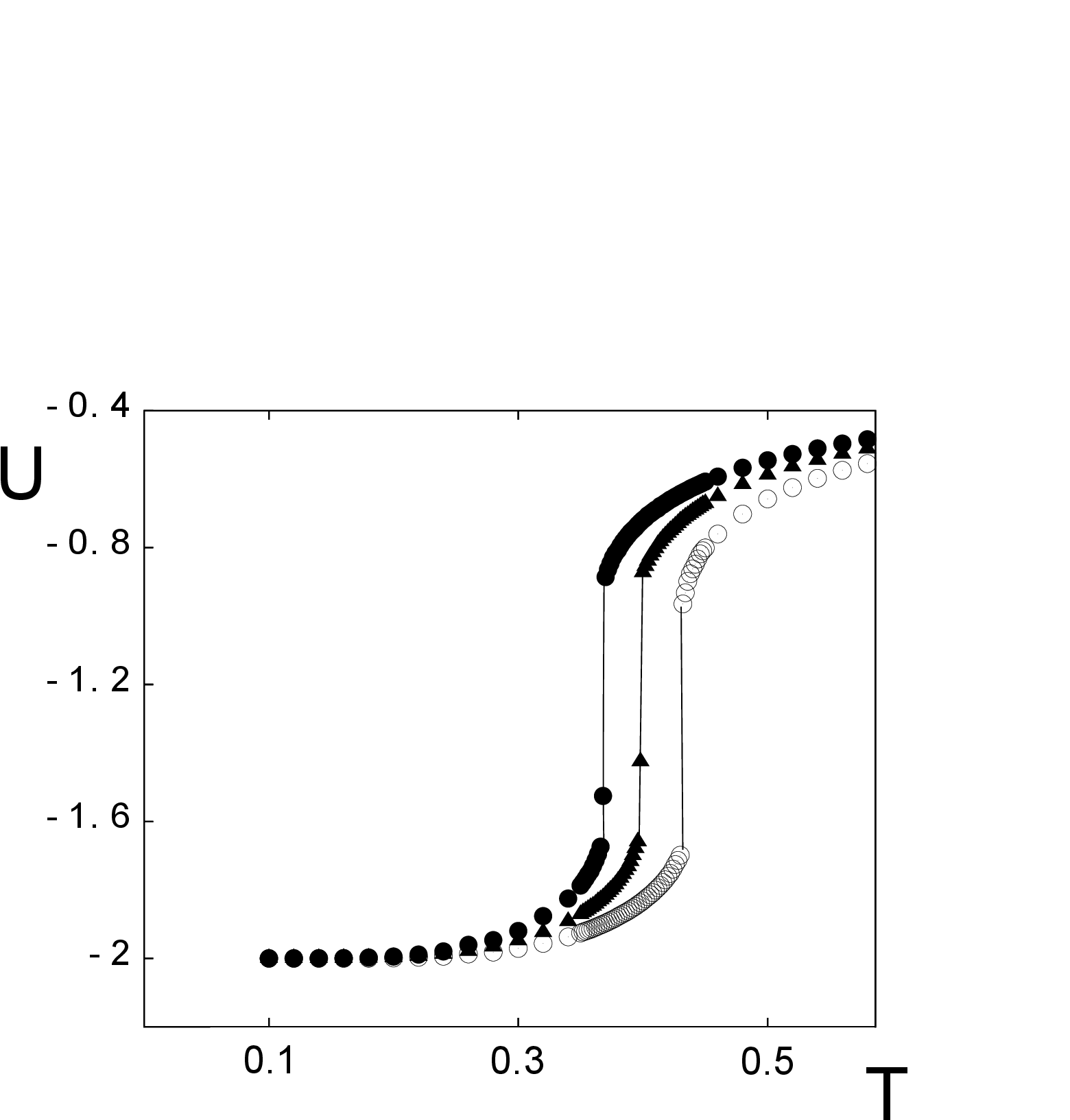}
\includegraphics[width=2.5in,angle=0]{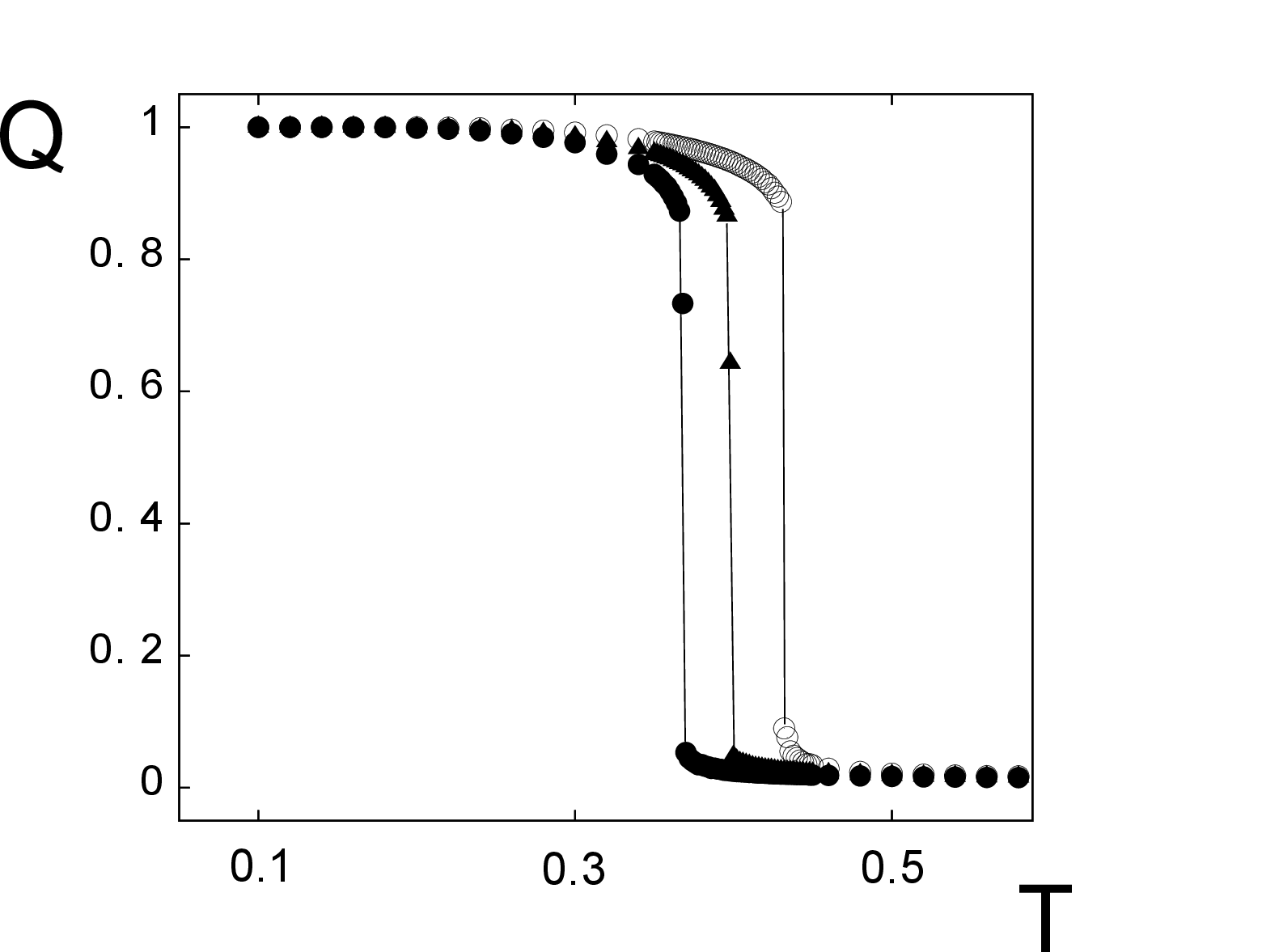}
\caption{Averaged  Potts energy $U$ and order parameter $Q$ vs temperature $T$ for $q=10$, $r=0.96$ (black circles), 1 (void circles), and 1.04 (black triangles),
with $N_x=N_y=100$ and $E_0=-0.5$. Lines are guides to the eye.} \label{fig:Eq10}
\end{figure}

We recall that the Potts model shows a first-order transition\cite{Wu} for $q>4$ at a finite temperature.
This is seen in Fig. \ref{fig:Eq10} where the averaged Potts energy $U$ defined by Eq. (\ref{u}) and the order parameter $Q$ are shown for three values of $r$.

Note that these quantities show a large discontinuity at the transition temperature.  These results confirm the first-order character of the transition.  Repeating the simulations for other values of $r$, we determine the Potts transition line in the space $(T,r)$ which agrees up to four digits with the exact critical line of Eq. (20), using $w_c = \sqrt {10}$.  Note that unlike in some first-order transitions, the slow heating and slow cooling of the system do not result in a hysteresis.  The energy barrier between the two phases is believed to be therefore not so high.

\begin{figure}
\centering
\includegraphics[width=2.5in,angle=0]{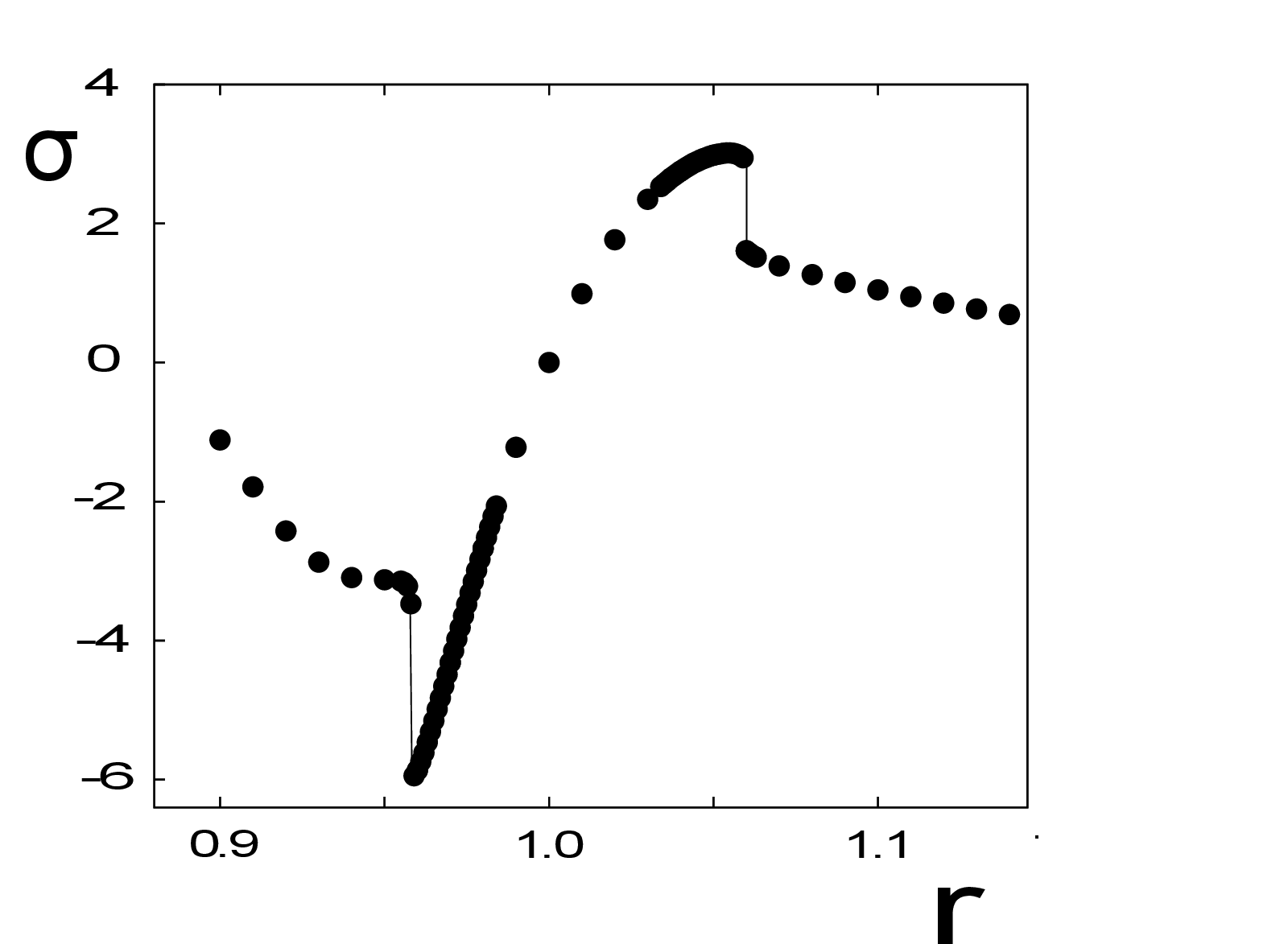}
\includegraphics[width=2.5in,angle=0]{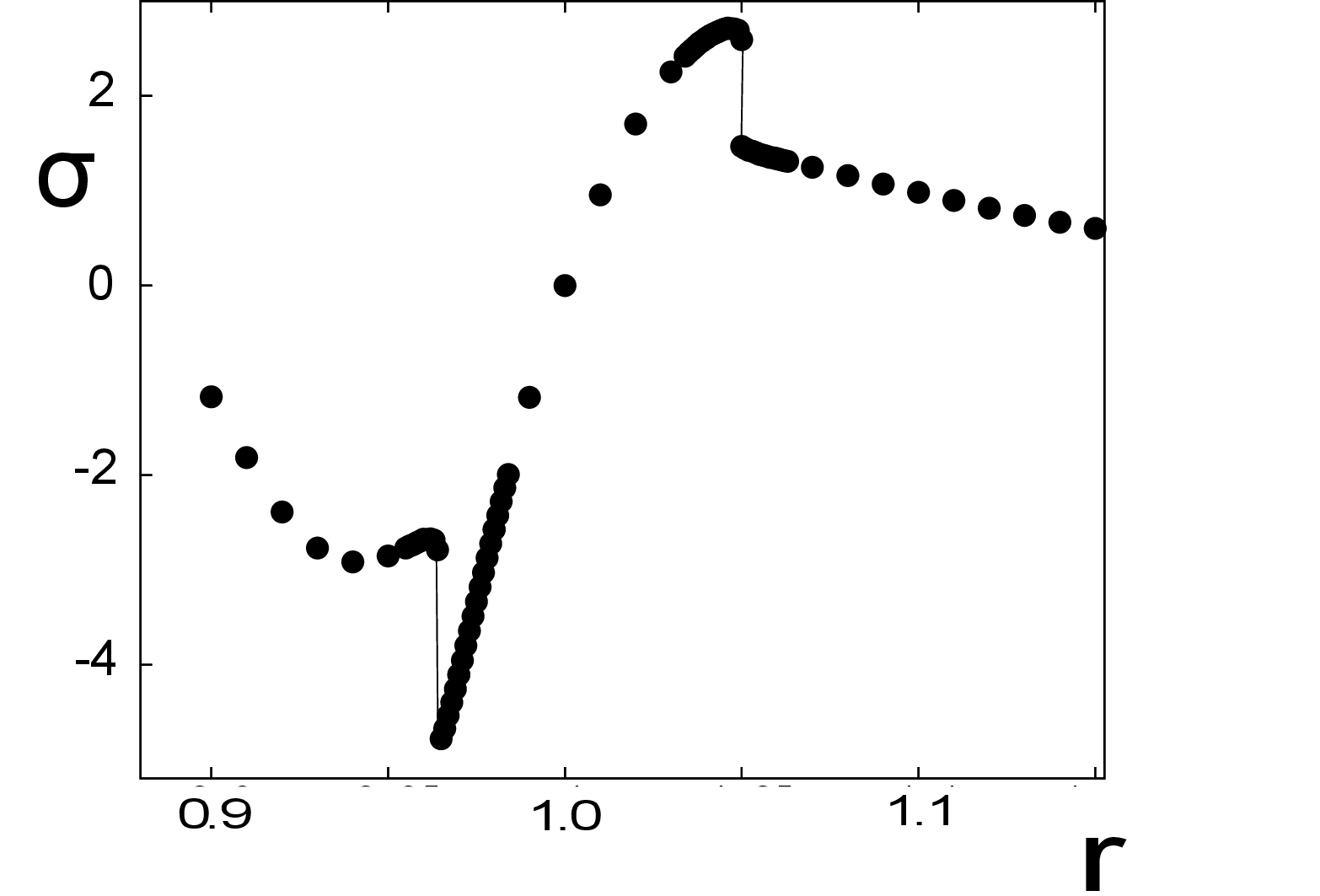}
\includegraphics[width=2.5in,angle=0]{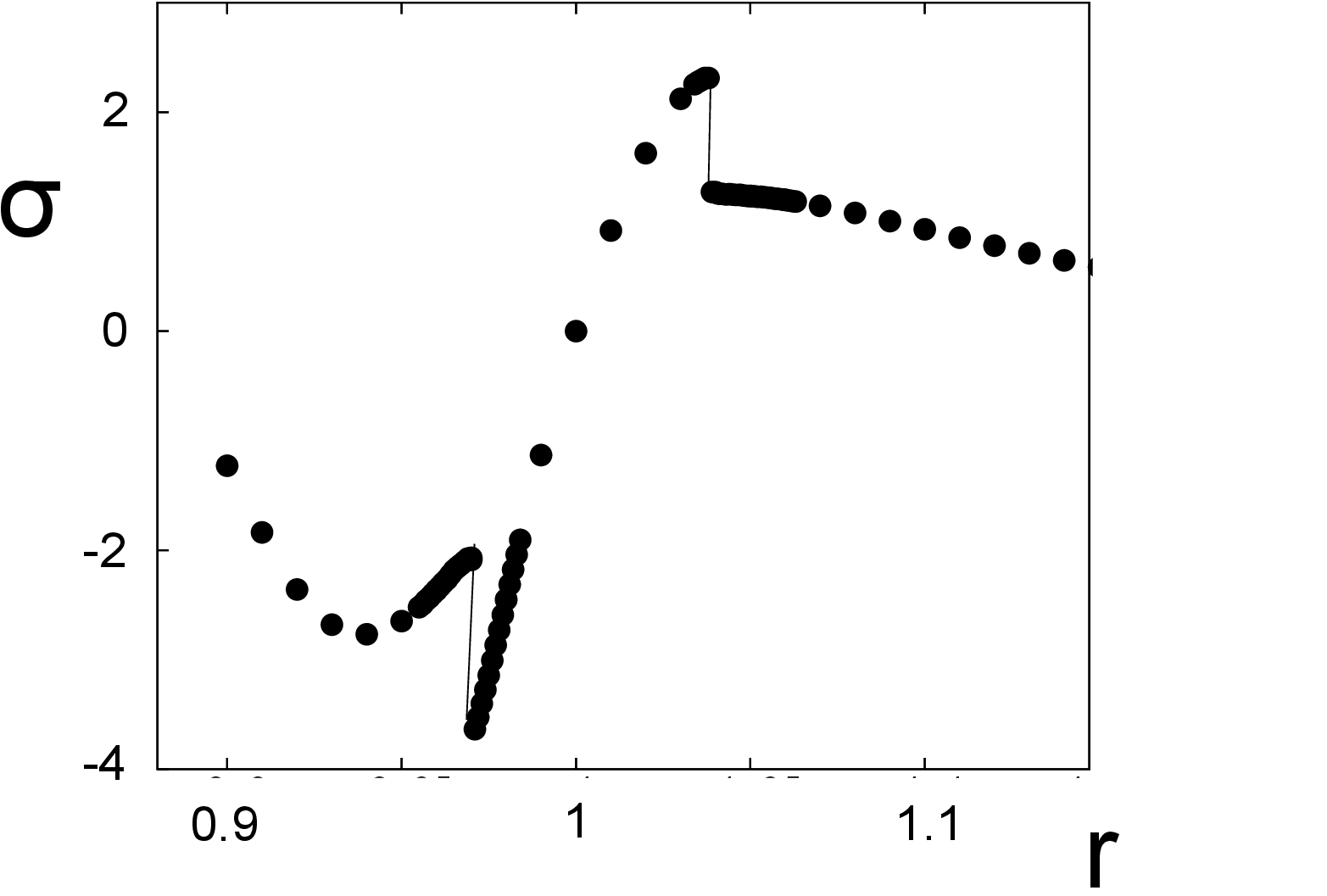}
\includegraphics[width=2.5in,angle=0]{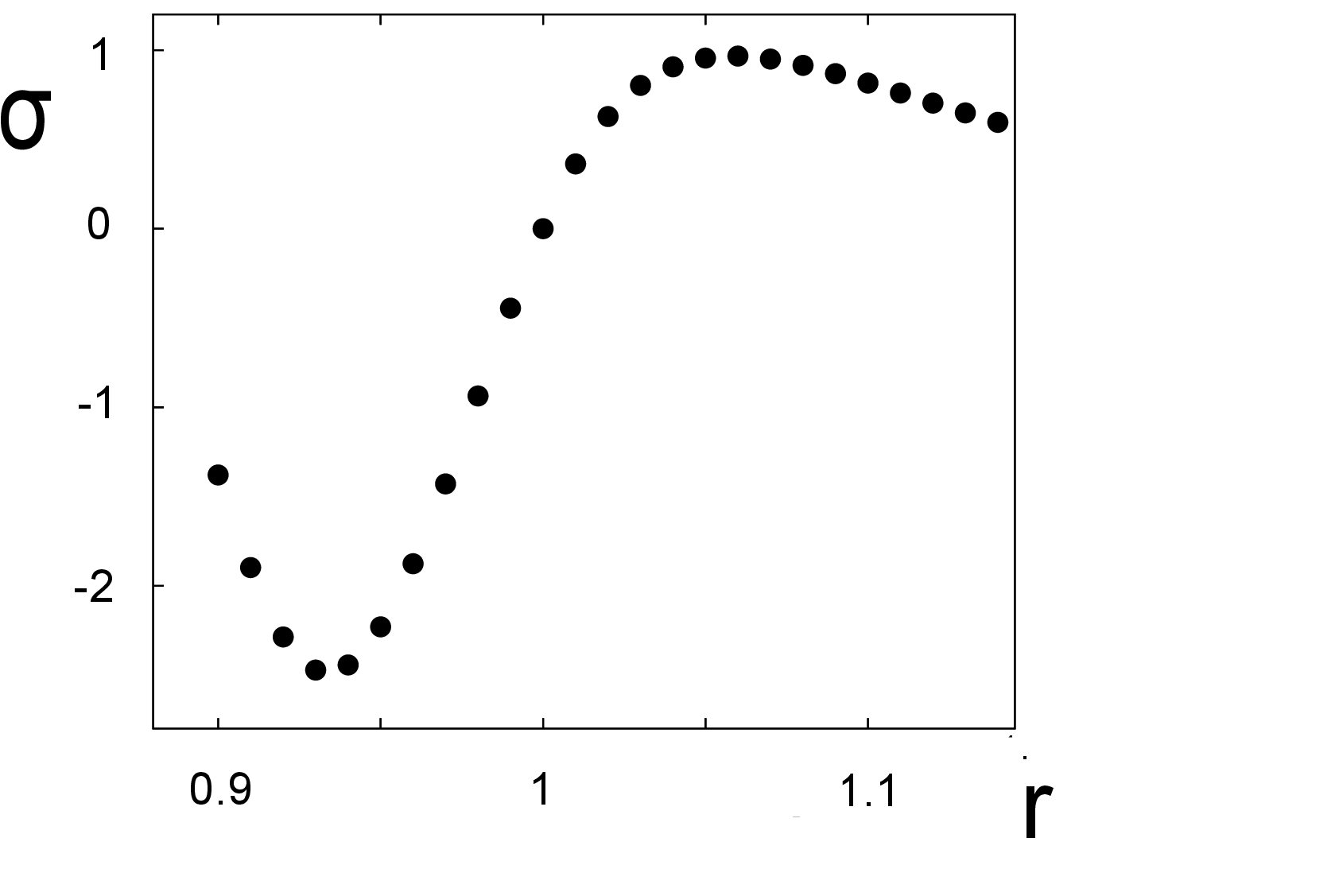}
\caption{Stress  vs  $r$ for $q=10$ at $T=0.36$ (upper left), 0.38 (upper right), 0.40 (lower left), 0.46 (lower right),
with $N_x=N_y=100$ and $E_0=-0.5$.  Lines are guides to the eye. See text for comments.} \label{fig:S1}
\end{figure}

We calculate the stress $\sigma$ using the Eq. (\ref{eqst1}) with $u$ obtained from MC simulations shown above in Fig. \ref{fig:Eq10}, $E(r)$ and $dE(r)/dr$ being given by Eqs. (\ref{LJ}) and (\ref{LJd}). This is done for many values of $r$ and $T$ around the Potts transition curve in search for unstable regions predicted by the RG analysis shown in Figs. \ref{fig:1}-\ref{fig:7}.  In practice, we fixed a temperature and then changed the value of $r$ across the Potts transition line by following a vertical line in Fig. \ref{fig:7}.  In doing so, we obtained for each $T$ the stress as a function of $r$.   At a given $T$, if there is no crossing of the Potts transition line then the stress behaves as shown in the curve for $T=0.46$ in Fig. \ref{fig:S1}:  $\sigma$ goes smoothly through a minimum at a compression position to  a maximum at a dilatation one.  The solid is stable in the region between the minimum and the maximum since the modulus $d\sigma/dr$ is positive.
On the other hand, when the system crosses the transition line by varying $r$, the stress exhibits jumps as seen in the three curves at $T=0.36$, 0.38, 0.40 in Fig. \ref{fig:S1}. The discontinuity in stress is due to the Potts energy discontinuity associated with the first-order transition of the 10-state Potts model.  The first-order nature in the case $q=10$ observed above enhances the instability.  As will be seen below, a qualitatively different instability occurs for $q=2,3,4$ where the transition is of second order.

\begin{figure}
\centering
\includegraphics[width=3.5in,angle=0]{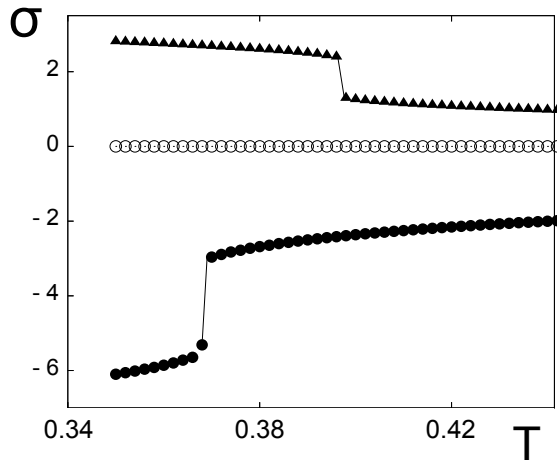}
\caption{Stress  vs  $T$ for $q=10$, at three values of $r$:  c1.04 (black triangles), 1 (void circles) and 0.96 (black circles),
with $N_x=N_y=100$ and $E_0=-0.5$. Note the discontinuities in the cases $r=0.96$ and 1.04.  Lines are guides to the eye.  See text for comments.} \label{fig:S-T}
\end{figure}

It is interesting to show now in Fig. \ref{fig:S-T} the stress  versus $T$ at a given $r$.  As seen the stress undergoes a discontinuity at the Potts transition temperature if $r\neq 1$, but the discontinuity vanishes for $r = 1$.

To close this sub-section, we emphasize that for the system studied here, the size effects are indistinguishable from $N_x=N_y=60$ up.

\subsection{Results for q = 2, 3, 4}

\begin{figure}
\centering
\includegraphics[width=3.5in,angle=0]{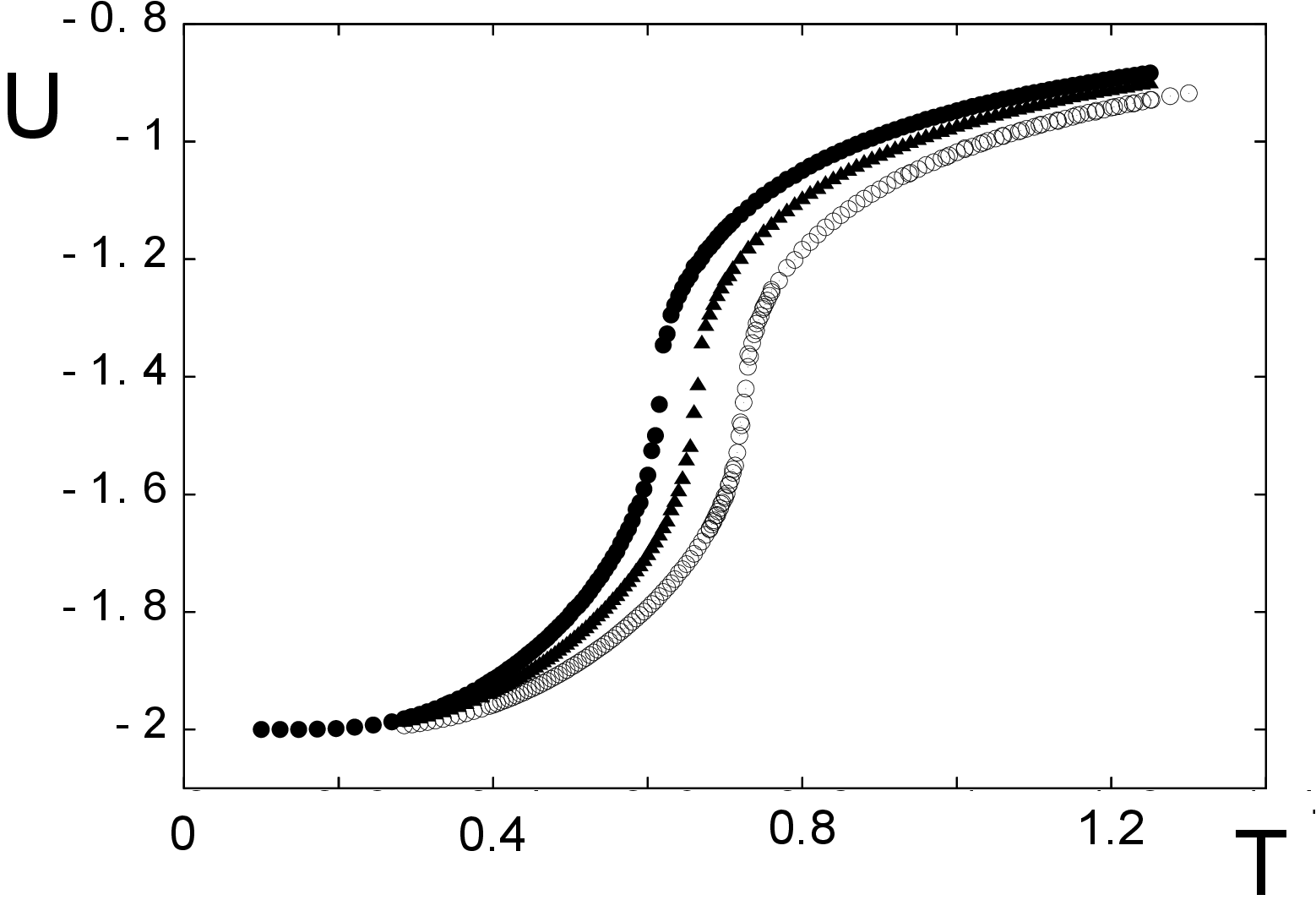}
\includegraphics[width=3.5in,angle=0]{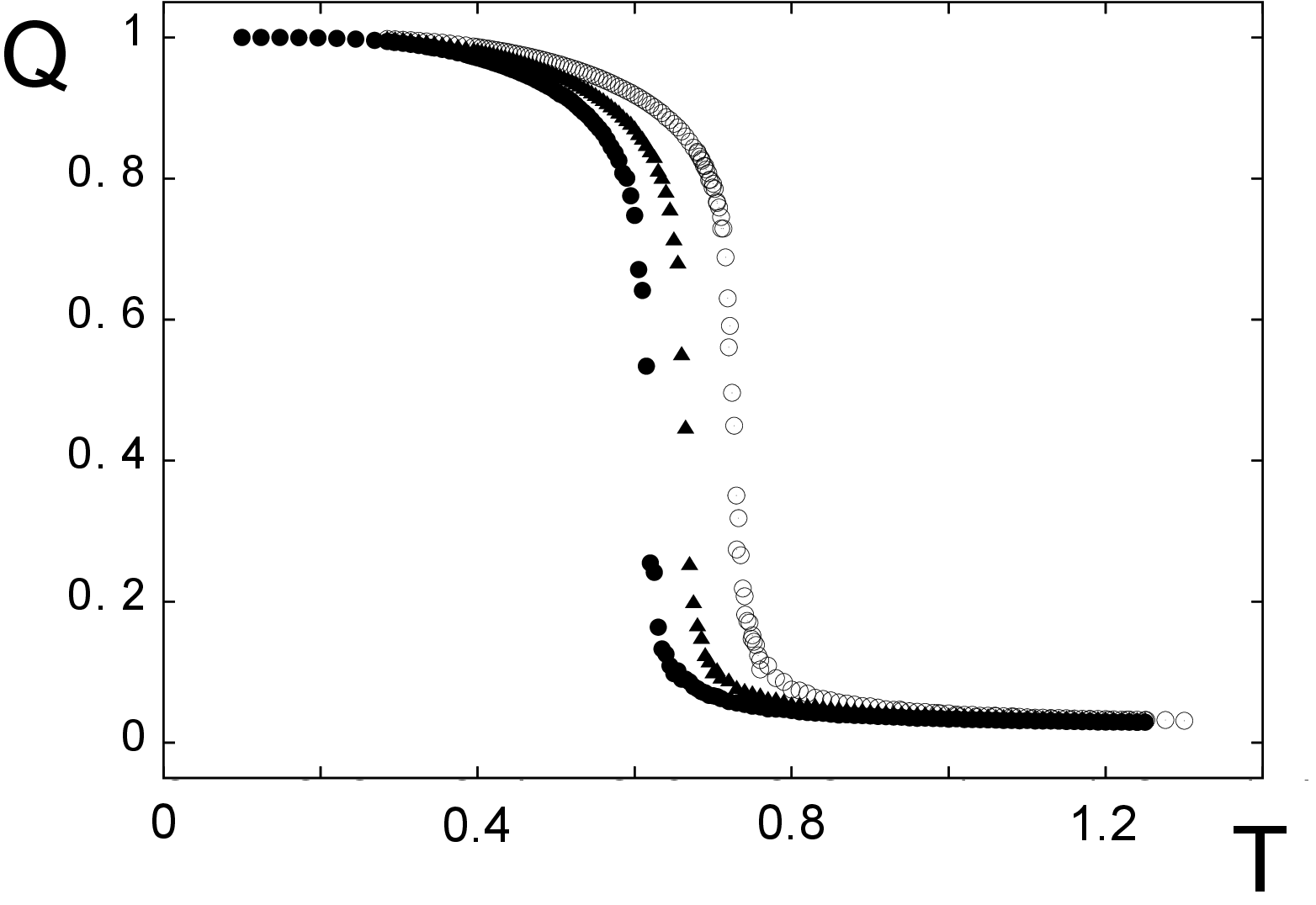}
\caption{Averaged  Potts energy $U$ and order parameter $Q$ vs temperature $T$ for $q=4$, $r=0.96$ (black circles), 1 (void circles), 1.04 (black triangles),
with $N_x=N_y=100$ and $E_0=-0.5$. } \label{fig:Eq4}
\end{figure}

\begin{figure}
\centering
\includegraphics[width=3.5in,angle=0]{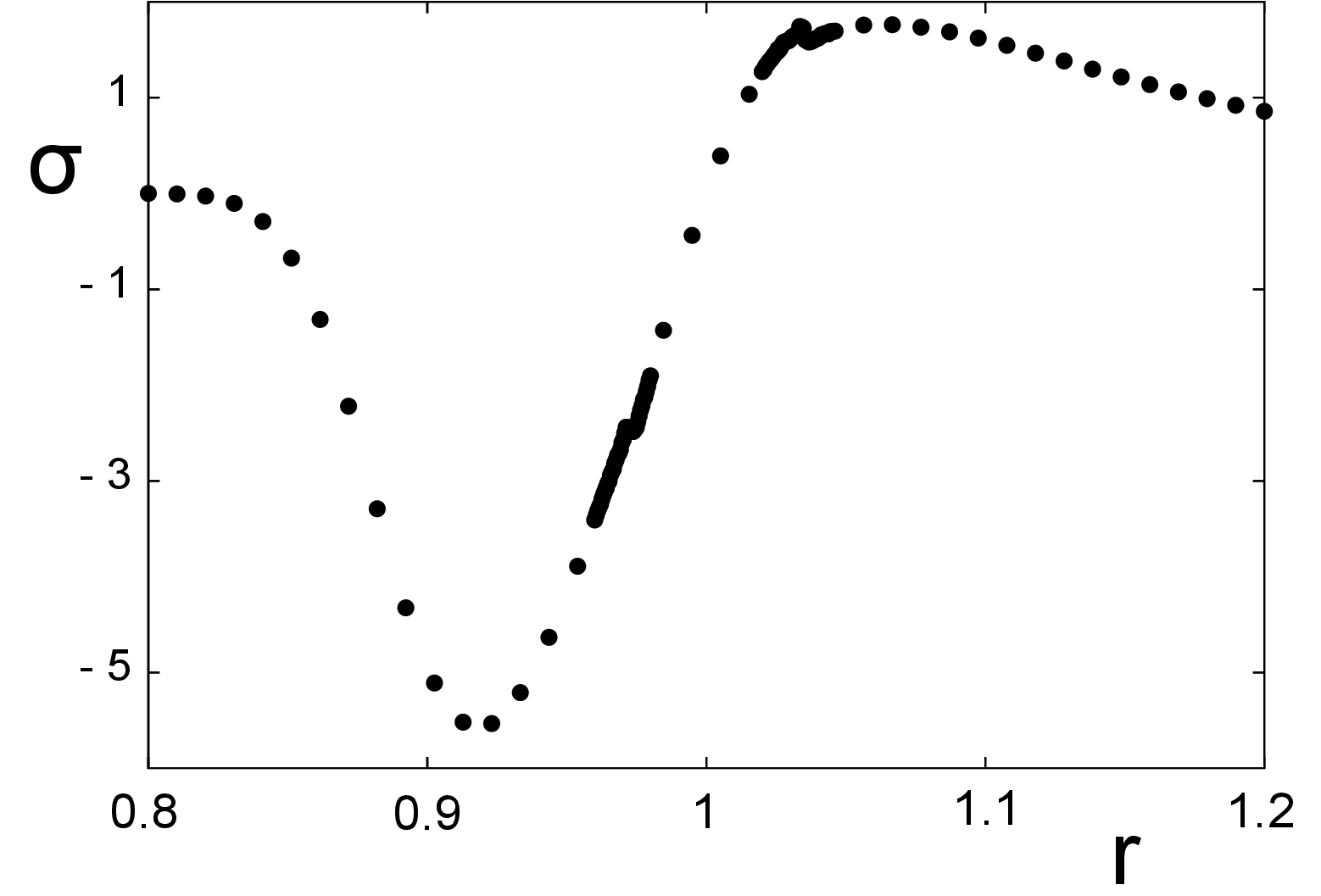}
\includegraphics[width=3.5in,angle=0]{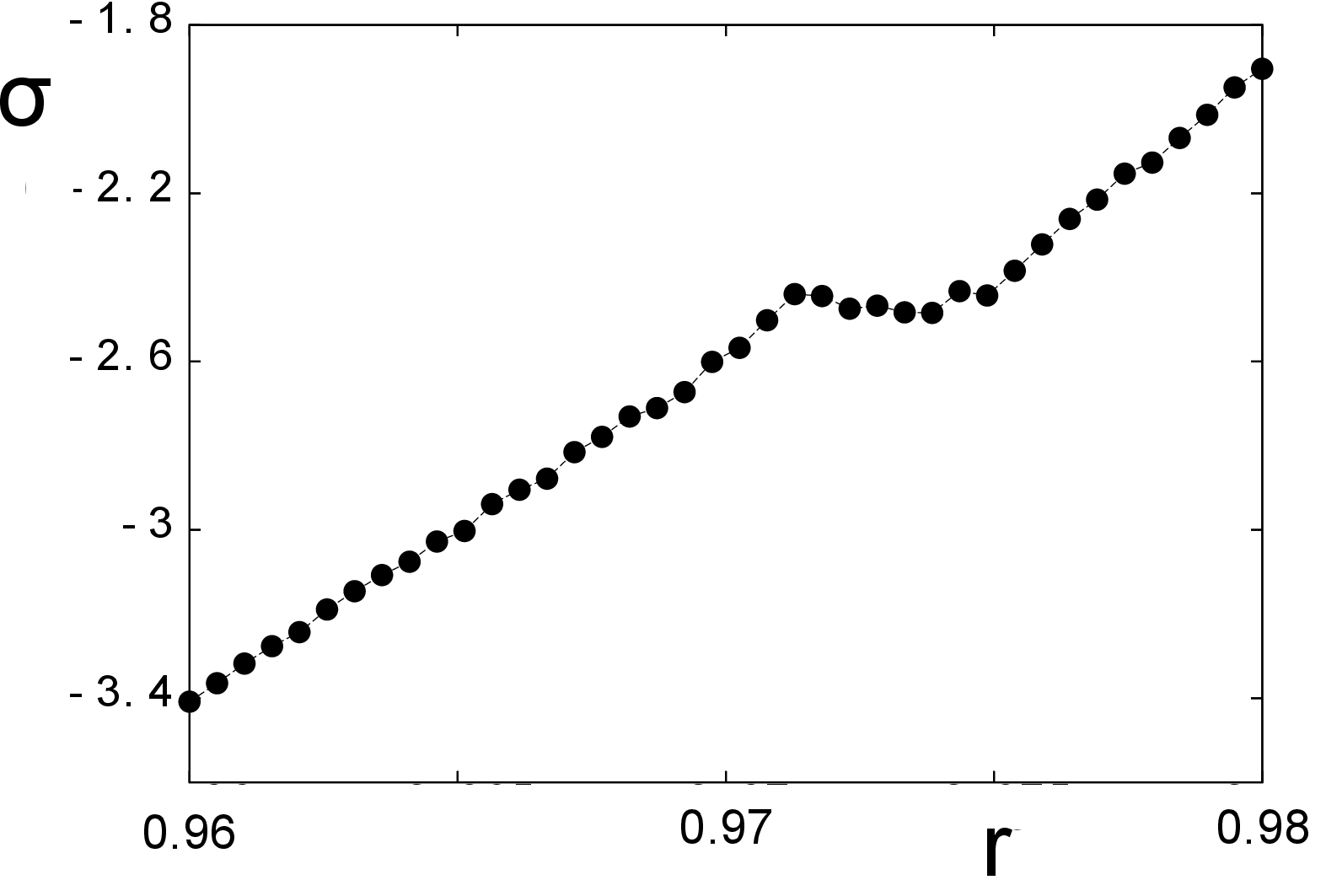}
\includegraphics[width=3.5in,angle=0]{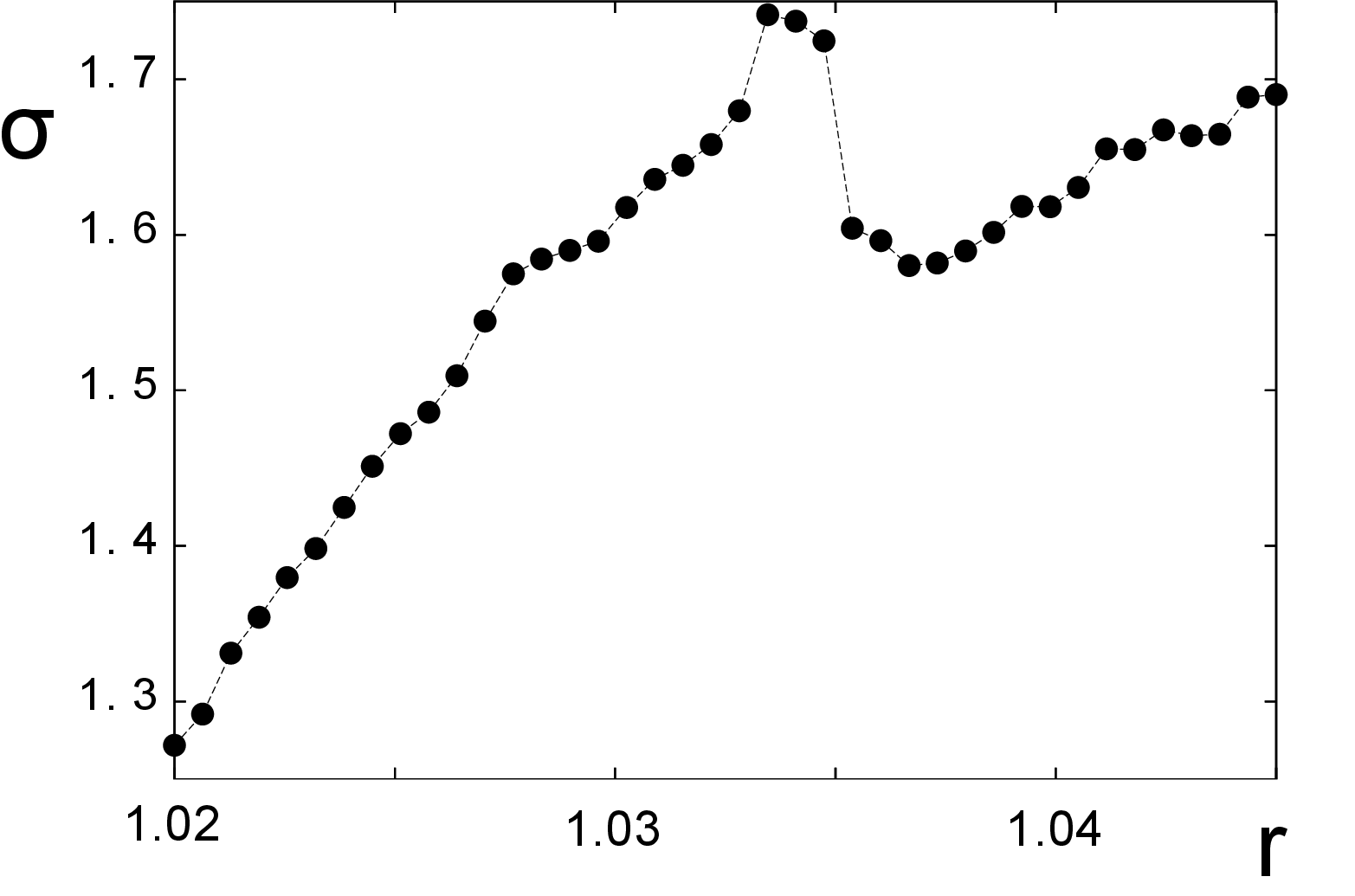}
\caption{Stress (upper curve) and zooms (lower curves) vs  $r$ for $q=4$, $T=0.67931$,
with $N_x=N_y=60$ and $E_0=-0.5$. } \label{fig:S}
\end{figure}

Let us consider the case where $q=4$. We show in Fig. \ref{fig:Eq4} the averaged Potts energy [Eq. (\ref{u})] and the order parameter $Q$ obtained by MC simulations as described above for $q=4$, $E_0=-0.5$ and three values of $r$.  The Potts transition is continuous with diverging heat capacity\cite{Wu}.  As a consequence of Eq. (10) the modulus becomes negative close to the transition temperature.  This instabilty implies the emergence of a van der Waals loop in the dependence of the stress on $r$.  We show this in Fig. \ref{fig:S}.  Hence the transition becomes weakly first order.

\begin{figure}
\centering
\includegraphics[width=3.5in,angle=0]{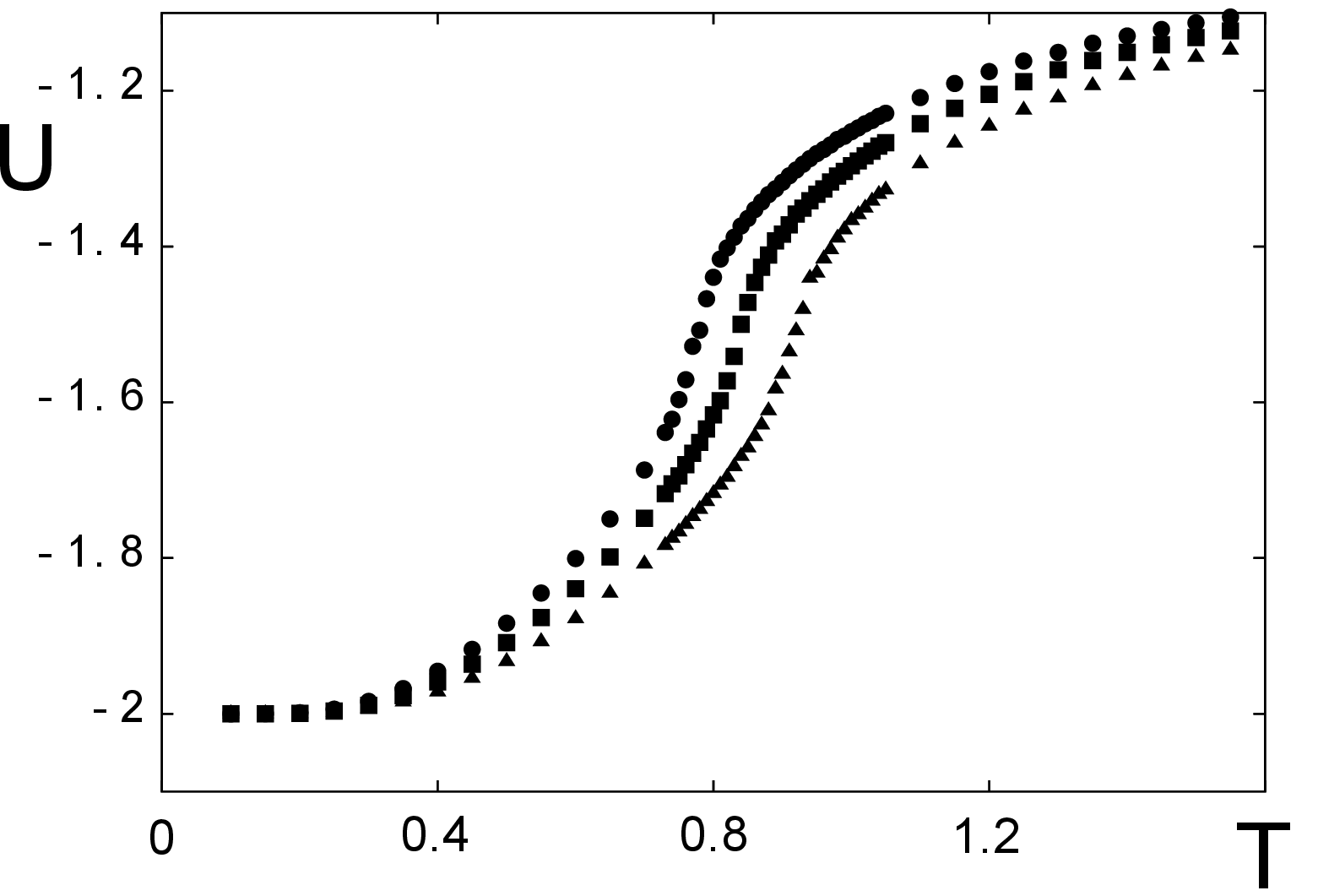}
\includegraphics[width=3.5in,angle=0]{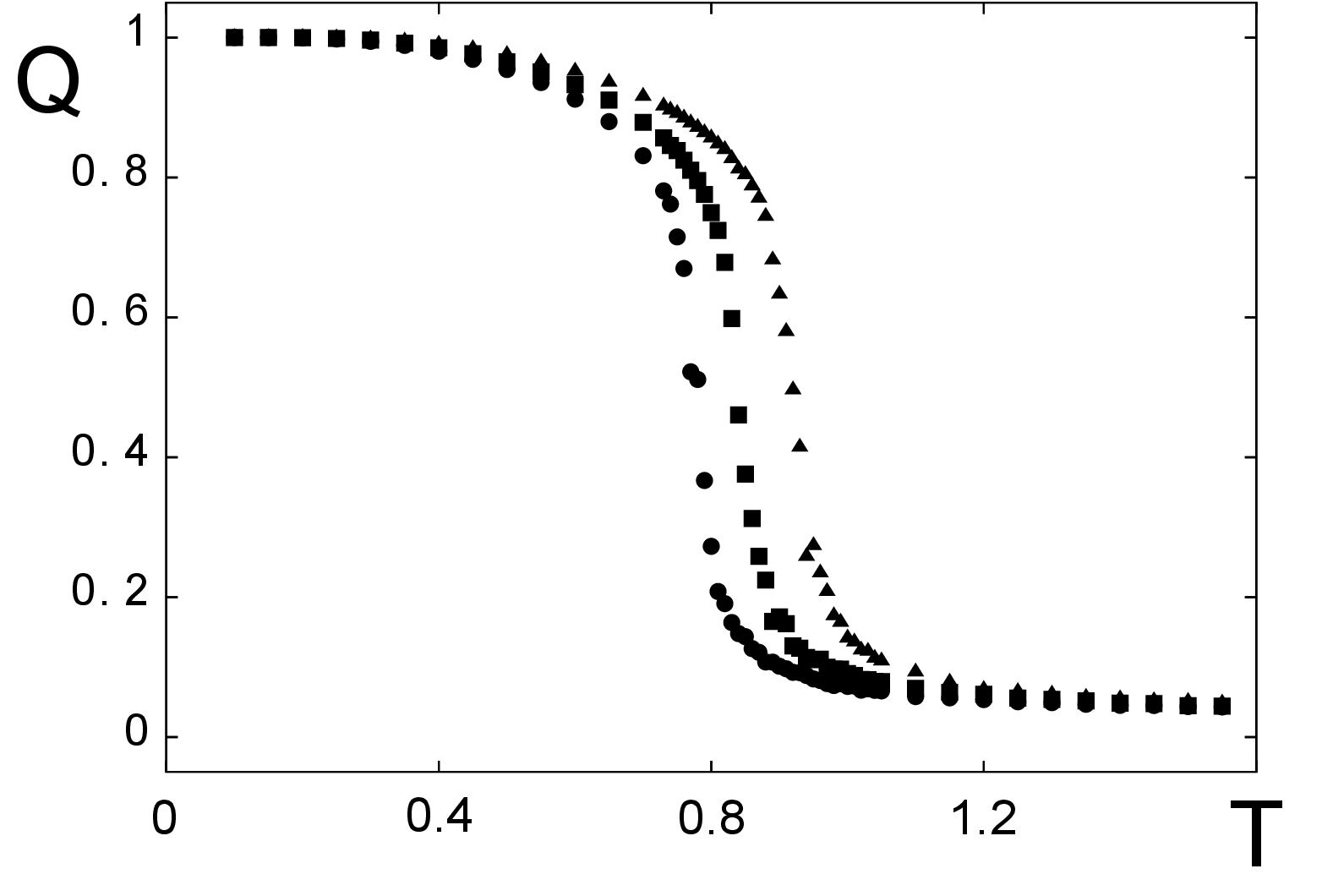}
\caption{Averaged  Potts energy $U$ and order parameter $Q$ vs temperature $T$ for $q=3$, $r=0.96$ (black circles), 1 (black triangles), 1.04 (black squares),
with $N_x=N_y=100$ and $E_0=-0.5$. } \label{fig:Eq3}
\end{figure}

\begin{figure}
\centering
\includegraphics[width=3.5in,angle=0]{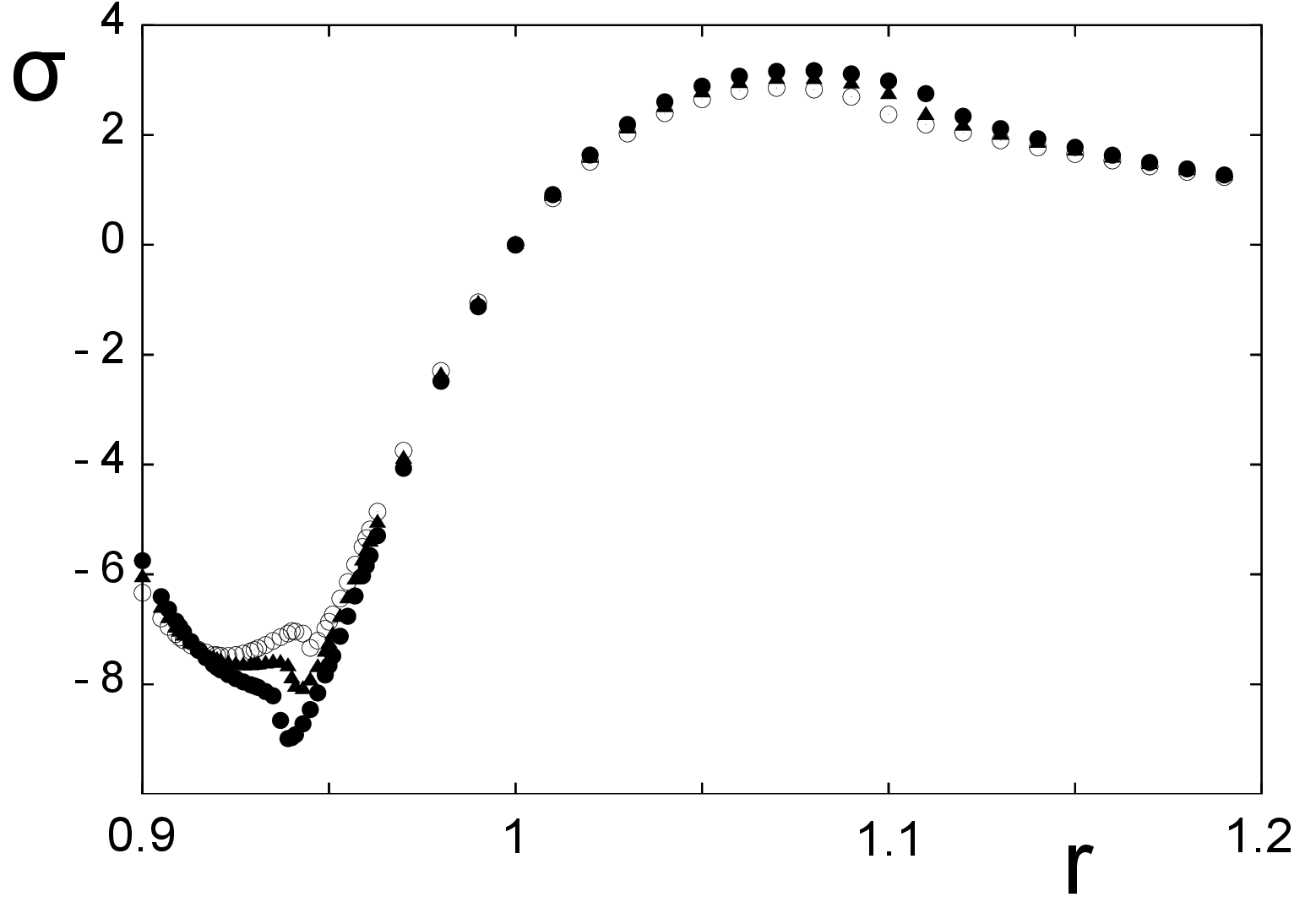}
\includegraphics[width=3.5in,angle=0]{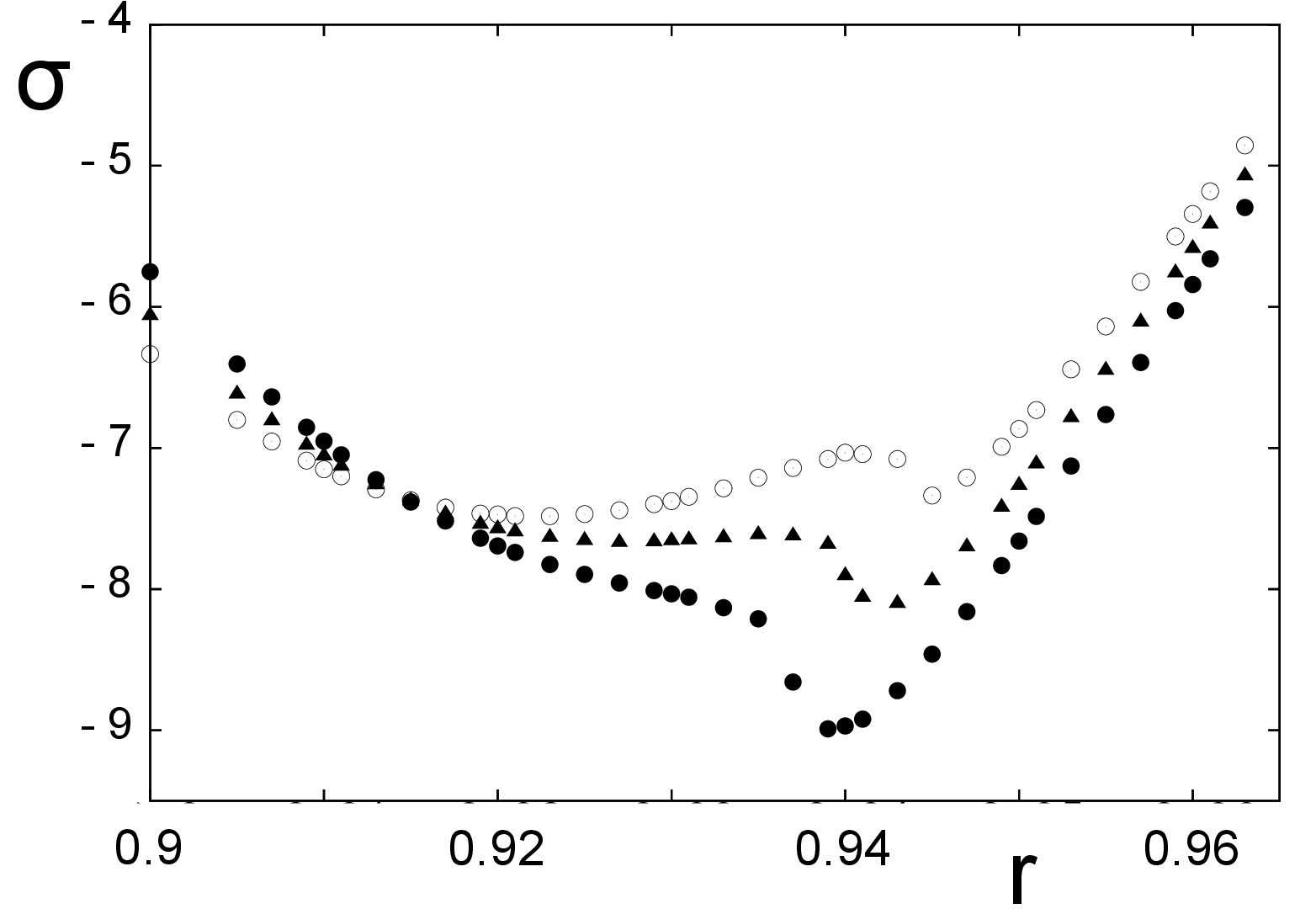}
\caption{Stress (upper) and zoom (lower) vs  $r$ for $q=3$, $T=0.5$ (black circles), 0.55 (black triangles) and 0.6 (void circles),
with $N_x=N_y=60$ and $E_0=-0.5$. } \label{fig:S3}
\end{figure}

\begin{figure}
\centering
\includegraphics[width=3.5in,angle=0]{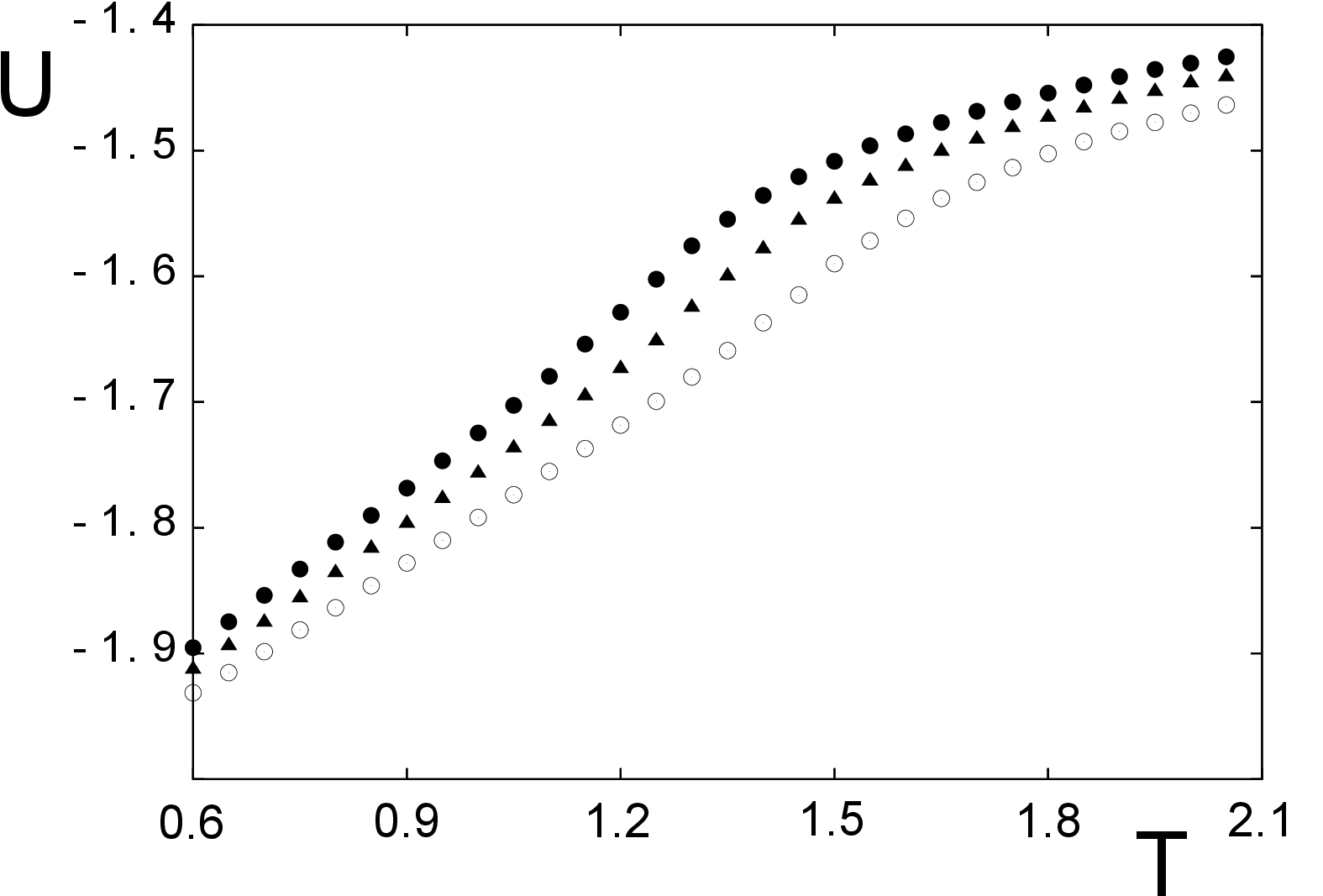}
\includegraphics[width=3.5in,angle=0]{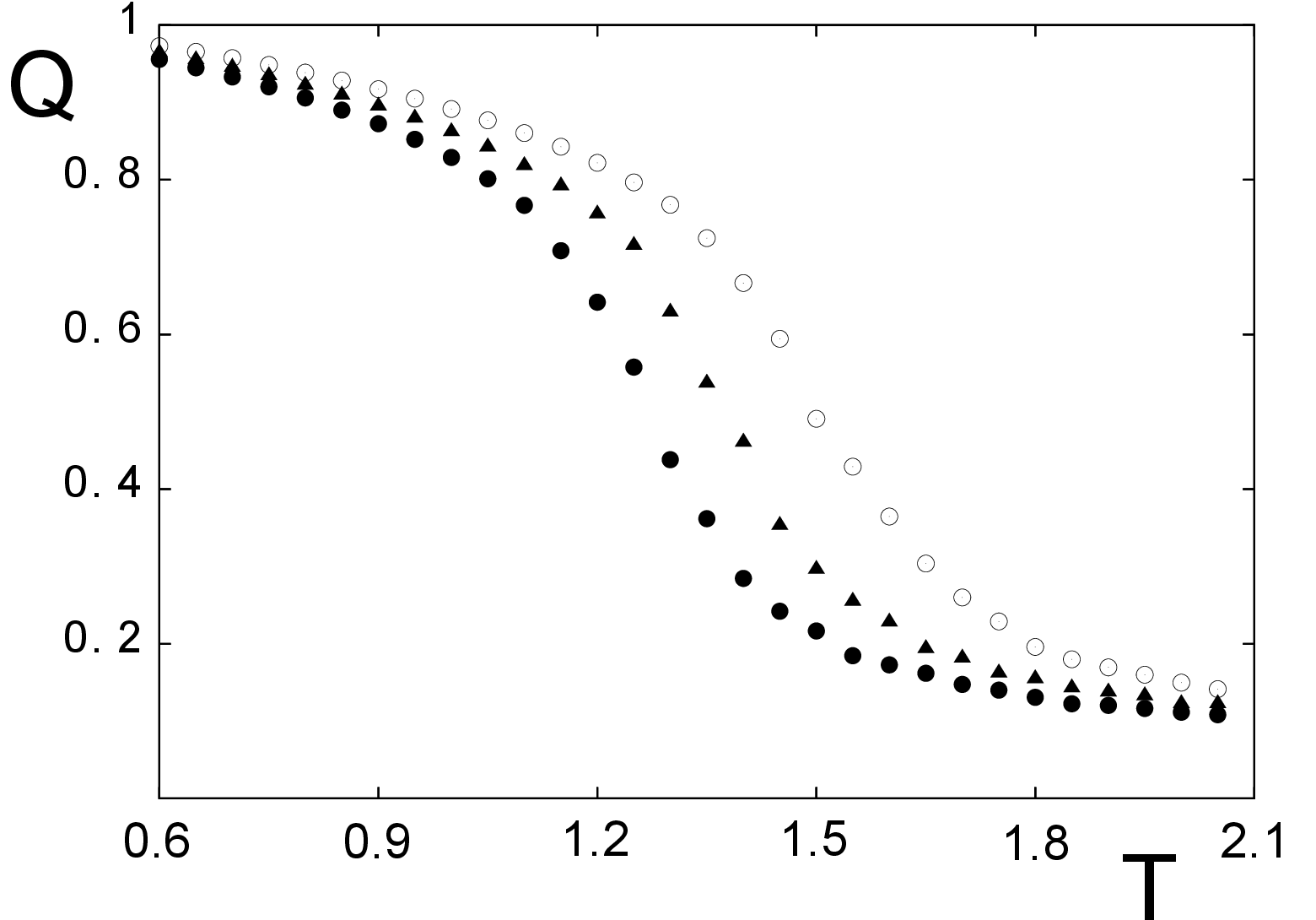}
\caption{Averaged Potts energy $U$ and order parameter $Q$ vs temperature $T$ for $q=2$, $r=0.96$ (black circles), 1 (void circles), 1.04 (black triangles),
with $N_x=N_y=100$ and $E_0=-0.5$. } \label{fig:Uq2}
\end{figure}

\begin{figure}
\centering
\includegraphics[width=3.5in,angle=0]{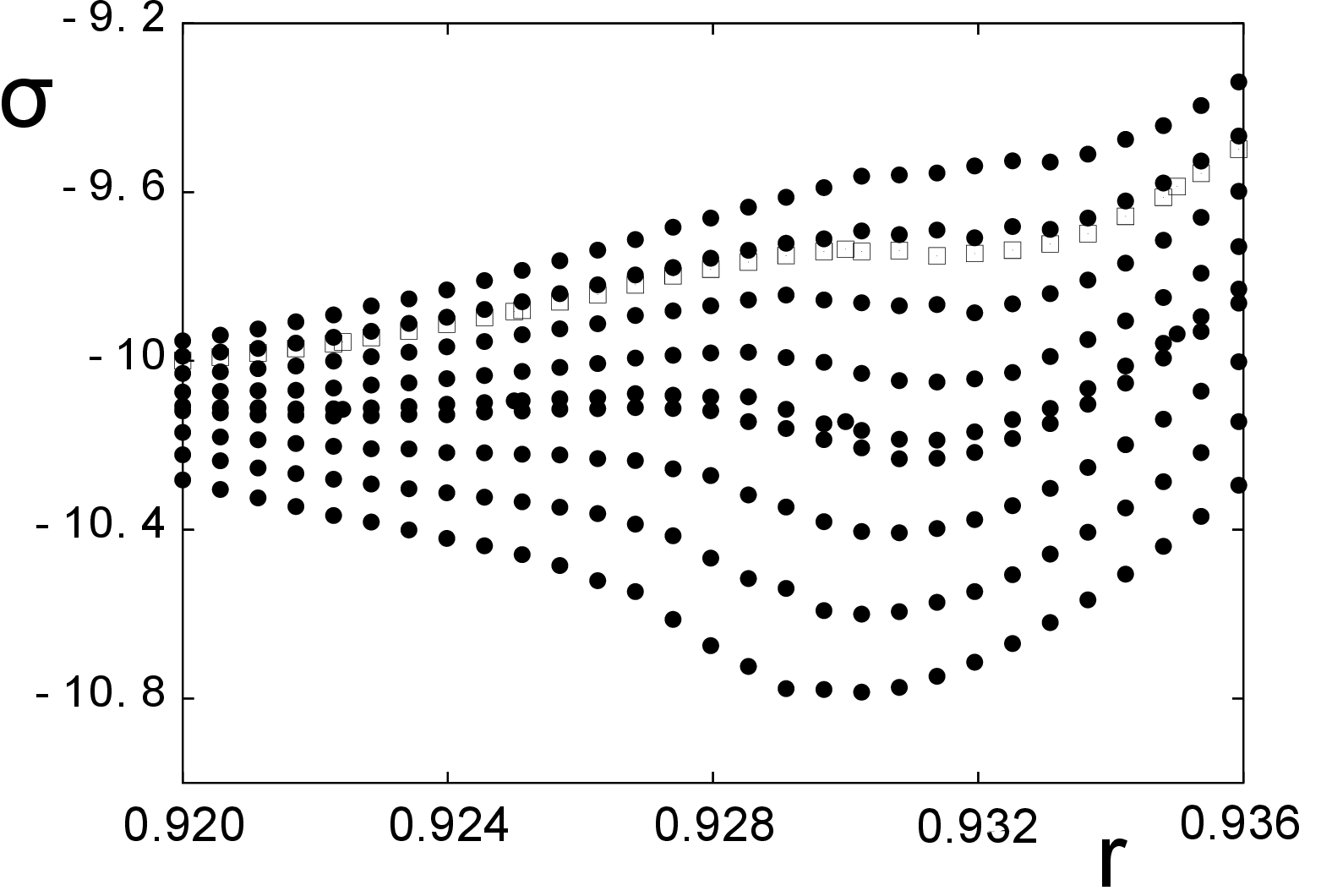}
\includegraphics[width=3.5in,angle=0]{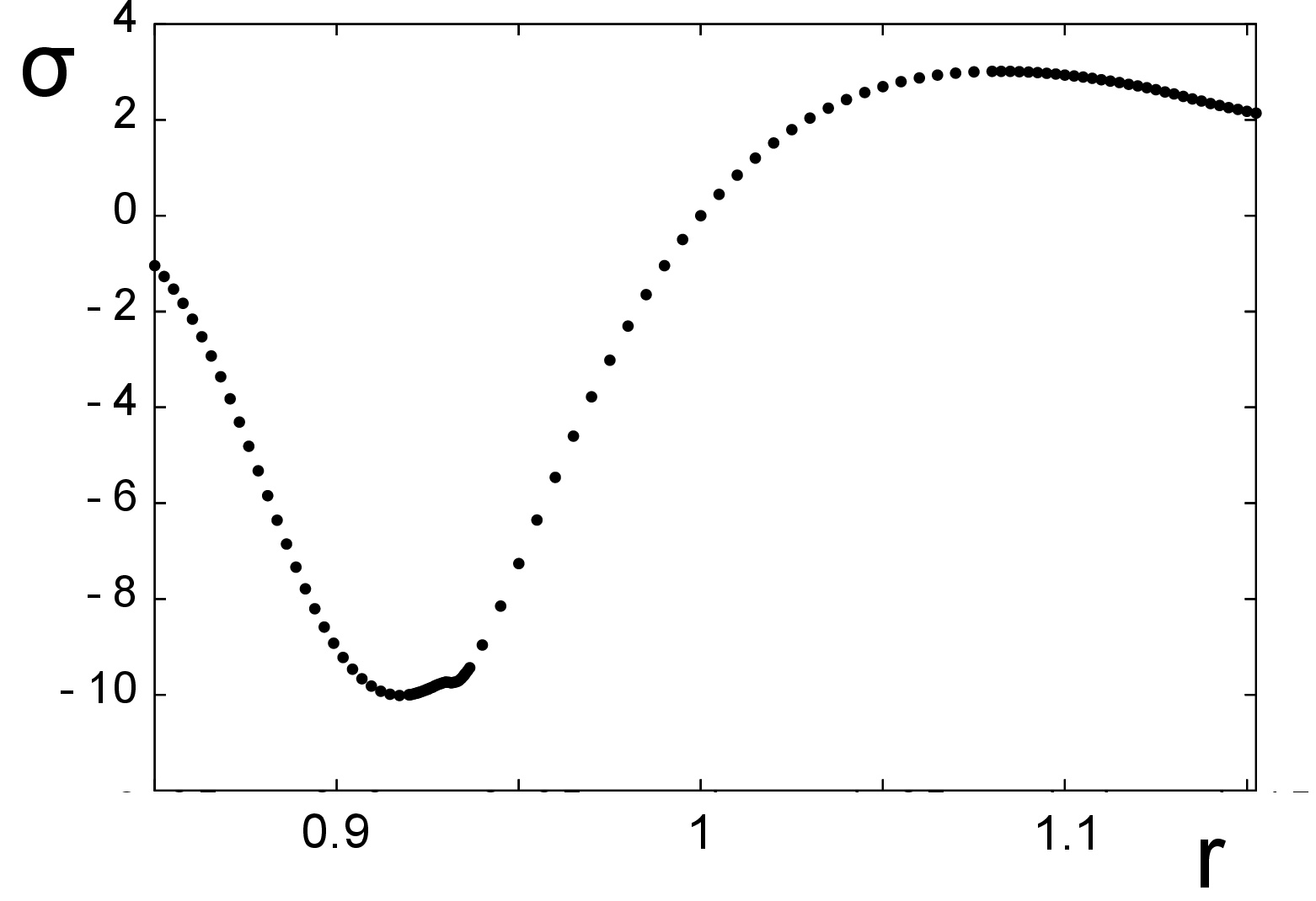}
\caption{Stress vs  $r$ near the instability region (upper) for $q=2$ with several temperatures from $T=0.535$ to 0.675: data points for $T=0.65$ are marked with void squares. This case is shown for a large region of $r$ in the lower figure.  $N_x=N_y=60$, $E_0=-0.5$.} \label{fig:S2}
\end{figure}

For $q=3$ and $q=2$, the results are shown in Figs. \ref{fig:Eq3}-\ref{fig:S3} and Figs. \ref{fig:Uq2}-\ref{fig:S2} respectively.  The replacement of the Potts continuous transition with a weakly first-order transition occurs here also because the Potts heat capacity is divergent for $q=2, 3, 4$.  As seen, the van der Waals loops, though weak, are present.

\section{ Conclusions}

In this paper, we have used renormalization group and Monte  Carlo simulations to study the phase diagram of a two-dimensional solid by using a model in which the interaction between neighboring atoms follows the Lennard-Jones potential.  We have mapped the model into a $q$-state Potts model and investigated the effect of both a uniform compression and a uniform expansion of the volume of the solid. In the temperature, inter-atomic distance plane we find a line of Potts transitions and stability boundaries where the stress as a function of inter-atomic distance has an extremum.  For the cases where the Potts heat capacity is divergent ($q = 2,3,4$ for the square lattice, $q > 6.8$ for the hierarchical diamond lattice) a van der Waals loop (instability) occurs close to the Potts transition and thus a weak first-order transition (using Maxwell construction) replaces the continuous transition. This is a remarkable result that warrants further analysis. Monte Carlo simulations for large $q$ values (where the square lattice Potts transition is discontinuous) indicate that the discontinuity in the Potts energy translates into a discontinuity in the stress as a function of the interatomic distance at the transition temperature. The influence of inter-atomic distance fluctuations on the nature of the phase transitions in this model will be addressed in a future study.  Finally, let us note that while it is known that the solid phase cannot survive at finite $T$ in two dimensions with continuous degrees of freedom of atom motions, our present study with discrete degrees of freedom shows some interesting behaviors which would serve as a starting point to study three-dimensional solids where melting mechanisms are not well understood.\cite{Gomez2001,Gomez2003}

M.K. wishes to thank the University of Cergy-Pontoise for hospitality while this work was carried out.

{}

\end{document}